\documentclass[draftclsnofoot, onecolumn, 12pt]{IEEEtran}
\usepackage{dsfont}
\usepackage{bbm}
\usepackage{hyperref} 
\usepackage{graphicx}
\usepackage{amssymb}
\usepackage[cmex10]{amsmath}
\DeclareMathOperator{\E}{\mathbb{E}}

\DeclareMathOperator{\Z}{\mathbb{Z}}
\DeclareMathOperator{\erf}{erf}

\newcommand{\Eb}[1]{{ \mathbb{E}\left[ #1 \right] }}
\newcommand{\Pb}[1]{{ \mathbb{P}\left[ #1 \right] }}
\newcommand{\1}[1]{\mathds{1}\left(#1\right)}

\newcommand {\Define} {\stackrel {\Delta} {=}  }

\newcommand{\mya}{\mathrel{\overset{\makebox[0pt]{{\tiny(a)}}}{=}}}

\newcommand{\myb}{\mathrel{\overset{\makebox[0pt]{{\tiny(b)}}}{=}}}

\newcommand{\tL}{\text{L}}
\newcommand{\tN}{\text{N}}
\newcommand{\mL}{m^{\left(\text{L}\right)}}
\newcommand{\mN}{m^{\left(\text{N}\right)}}
\newcommand{\nL}{n^{\left(\text{L}\right)}}
\newcommand{\nN}{n^{\left(\text{N}\right)}}

\newcommand{\aL}{\alpha^{(\tL)}}
\newcommand{\aN}{\alpha^{(\tN)}}
\newcommand{\ack}{\alpha^{(c)}}
\newcommand{\aco}{\alpha^{\ncob}}

\newcommand{\nco}{{c}_o}

\newcommand{\nc}{{c}}
\newcommand{\ncob}{\left({c}_o\right)}
\newcommand{\ncb}{\left({c}\right)}
   
\newcommand{\ncp}{\left({c^{\prime}}\right)}

\newcommand{\gleq}{{{{k_0(\bm \varrho) = \lfloor \zeta(\bm \varrho) \rfloor}\atop \geq}\atop <}}

\usepackage{bm}
\usepackage{mathtools}
\usepackage{epsfig,makeidx,color}
\usepackage{cite}
\usepackage{stfloats}
\usepackage{subfigure}
\usepackage{psfrag}
\usepackage[mathscr]{euscript}
\usepackage{chngcntr}
\usepackage{lineno}
\usepackage[T1]{fontenc}
\usepackage{microtype}
\usepackage{wasysym}
\usepackage{footnote}
\usepackage{acronym}  
\usepackage{booktabs} 
\usepackage[table]{xcolor}
\usepackage[table]{xcolor}
\usepackage{amsbsy}
\usepackage{amssymb}
\usepackage{psfrag}
\usepackage[mathscr]{euscript}
\usepackage{lipsum}
\usepackage{cite}
\usepackage{nicefrac}
\usepackage[english]{babel}
    \usepackage[font=small,labelsep=space]{caption}
\makeatletter
\def\citenoauxwrite#1{\begingroup
\@fileswfalse
\cite{#1}\relax
\endgroup}
\makeatother

\acrodef{CCDF}{complementary cumulative distribution function}
\acrodef{CF}{characteristic function}
\acrodef{PPP}{Poisson point process}
\acrodef{CSI}{channel state information}
\acrodef{OFDM}{orthogonal frequency division multiplexing}
\acrodef{OFDMA}{orthogonal frequency division multiple access}

\acrodef{RV}{random variable}
\acrodef{i.i.d.}{independent, identically distributed}
\acrodef{PMF}{probability mass function}
\acrodef{PDF}{probability distribution function}
\acrodef{CDF}{cumulative distribution function}
\acrodef{ch.f.}{characteristic function}
\acrodef{AWGN}{additive white Gaussian noise}
\acrodef{SNR}{signal-to-noise ratio}
\acrodef{LRT}{likelihood ratio test}
\acrodef{DRT}{distance ratio test}
\acrodef{GLRT}{generalized likelihood ratio test}
\acrodef{CRLB}{Cram\'{e}r-Rao lower bound}
\acrodef{CRB}{Cram\'{e}r-Rao bound}
\acrodef{ZZLB}{Ziv-Zakai lower bound}
\acrodef{ZZB}{Ziv-Zakai bound}
\acrodef{LOS}{line-of-sight}
\acrodef{ToF}{time-of-flight}
\acrodef{NLOS}{non-line-of-sight}
\acrodef{GDOP}{geometric dilution of precision}
\acrodef{GPS}{Global Positioning System}
\acrodef{FIM}{Fisher information matrix}
\acrodef{PEB}{position error bound}
\acrodef{SPEB}{squared position error bound}
\acrodef{TOA}{time-of-arrival}
\acrodef{TOF}{time-of-flight}
\acrodef{WSN}{wireless sensor network}
\acrodef{MAC}{medium access control}
\acrodef{RSS}{received signal strength}
\acrodef{WAF}{wall attenuation factor}
\acrodef{TDOA}{time difference-of-arrival}
\acrodef{RF}{radiofrequency}
\acrodef{RTT}{round-trip time}
\acrodef{AOA}{angle-of-arrival}
\acrodef{MF}{matched filter}
\acrodef{ED}{energy detector}
\acrodef{ML}{maximum likelihood}
\acrodef{MSE}{mean-square error}
\acrodef{RMSE}{root-mean-square error}
\acrodef{LEO}{localization error outage}
\acrodef{ppm}{part-per-million}
\acrodef{ACK}{acknowledge}
\acrodef{UWB}{Ultrawide bandwidth}
\acrodef{TNR}{threshold-to-noise ratio}
\acrodef{LS}{least squares}
\acrodef{IR-UWB}{impulse radio UWB}
\acrodef{FCC}{Federal Communications Commission}
\acrodef{TH}{time-hopping}
\acrodef{PPM}{pulse position modulation}
\acrodef{MUI}{multi-user interference}
\acrodef{PDP}{power delay profile}
\acrodef{BPZF}{band-pass zonal filter}
\acrodef{SIR}{signal-to-interference ratio}
\acrodef{RFID}{radio frequency identification}
\acrodef{WPAN}{wireless personal area network}
\acrodef{WWB}{Weiss-Weinstein bound}
\acrodef{DP}{direct path}
\acrodef{MF}{matched filter}
\acrodef{MMSE}{minimum-mean-square-error}
\acrodef{SBS}{serial backward search}
\acrodef{SBSMC}{serial backward search for multiple clusters}
\acrodef{NBI}{narrowband interference}
\acrodef{WBI}{wideband interference}
\acrodef{INR}{interference-to-noise ratio}
\acrodef{CR}{channel response}
\acrodef{CIR}{channel impulse response}
\acrodef{CR}{channel  response}
\acrodef{RADAR}{radar}
\acrodef{MUR}{Multistatic radar}
\acrodef{JBSF}{jump back and search forward}
\acrodef{HDSA}{high-definition situation-aware}
\acrodef{RRC}{root raised cosine}
\acrodef{ST}{simple thresholding}
\acrodef{BTB}{Bellini-Tartara bound}
\acrodef{P-Max}{$P$-Max}  
\acrodef{MIMO}{multiple-input multiple-output}
\acrodef{MAP}{maximum a posteriori}
\acrodef{FG}{factor graph}
\acrodef{OP}{outage probability}
\acrodef{WED}{wall extra delay}
\acrodef{RMS}{root mean square}
\acrodef{SPAWN}{sum-product algorithm over a wireless network}
\acrodef{MDD}{minimum distance distribution}
\acrodef{MAP}{maximum a posteriori probability}
\acrodef{PAR}{probabilistic association rule}

\usepackage{color}
\usepackage{dsfont}
\usepackage{bbm}

\newcommand{\Ws}[2]{{W_{}^{}}} 

\newcommand{\TSIR}[2]{{\tau_{}^{}}}

\DeclareMathAlphabet{\mathsf}{OML}{cmbr}{m}{it}

\newtheorem{definition}{Definition}
\newtheorem{theorem}{Theorem}
\newtheorem{lemma}{Lemma}
\newtheorem{corollary}{Corollary}
\newtheorem{proposition}{Proposition}

\newtheorem{remark}{Remark}

\newcommand{\HGF}[3]{{}_{#1}F_{#2}\!\left(#3\right)}

\newcommand{\bd}{\begin{description}}
\newcommand{\ed}{\end{description}}
\newcommand{\be}{\begin{enumerate}}
\newcommand{\ee}{\end{enumerate}}
\newcommand{\bi}{\begin{itemize}}
\newcommand{\ei}{\end{itemize}}
\newcommand{\bl}{\begin{list}}
\newcommand{\el}{\end{list}}
\newcommand{\bt}{\begin{tabbing}}
\newcommand{\et}{\end{tabbing}}

\acrodef{BS}{base station}
\acrodef{AP}{access point}
\acrodef{TDD}{time-division duplexing}

\begin{document}

\newcommand{\paperTitle}{Base Station Coordination Scheme for Multi-tier Ultra-dense Networks}
%
%
 
 

\title{\paperTitle}

\author{
        Sudarshan~Mukherjee, \textit{Member, IEEE},
        Dongsun Kim, \textit{Student Member, IEEE}, and
        Jemin~Lee, \textit{Member, IEEE}
%
%
%
\thanks{
        S. Mukherjee was with the Department of Information \& Communication Engineering (ICE), DGIST, Republic of Korea. He is currently with the Department of Electronics \& Electrical Engineering, IIT Guwahati, India (e-mail:\texttt{smukherjee@iitg.ac.in}). 
}
\thanks{
        D. Kim and J. Lee are with    
       the Department of Information \& Communication Engineering (ICE), DGIST, Republic of Korea (e-mail:\texttt{yidaever@dgist.ac.kr, jmnlee@dgist.ac.kr}). 
}

   \thanks{A part of the material presented in this paper was presented in IEEE Wireless Communications and Networking Conference (WCNC), 2020 \cite{WCNC20}.}
   
   \thanks{
       The corresponding author is J. Lee. 
       }
       
    \thanks{
        This work was supported by Institute for Information \& communication Technology Promotion(IITP) grant funded by the Korea government(MSIT) (No. 2018-0-01410, Development of Radio Transmission Technologies for High Capacity and Low Cost in Ultra Dense networks).
}
    
}

\maketitle 

%

%

%
\setcounter{page}{1}
\acresetall

\vspace{-2 cm}

\begin{abstract}
In this paper, we consider a relative received link power (RRLP)-based coordinated multi-point (CoMP) joint transmission (JT) in the multi-tier ultra-dense networks (UDN). In this CoMP scheme, we identify the cooperating base stations (BSs) by comparing the average received link power (ARLP) of the neighbouring BSs with respect to the BS having the strongest ARLP (i.e., the main link BS) to a user. To analyze the performance of this CoMP scheme in the downlink multi-tier UDN, we first approximate the received signal power distribution, and derive the coverage probability using stochastic geometry. After revisiting the area spectral efficiency (ASE) to make it more suitable for CoMP transmission in UDN, we also analyze the ASE and the network energy efficiency (NEE). Using simulations, we validate the derived coverage probability, and investigate the CoMP performance in multi-tier UDN. Our simulations show that the RRLP-based CoMP scheme can outperform the fixed number of strongest BS-based CoMP scheme in the high BS density regime. Our study of the NEE performance reveals that not only the RRLP-based CoMP scheme is more efficient than conventional non-CoMP transmission scenario, but also its NEE performance improves with the average number of cooperating BSs. 
\end{abstract}



\acresetall


\vspace{-0.4 cm}

\section{Introduction}
In recent years, development of the fifth generation (5G) new radio (NR) has been identified to be contingent upon a diverse array of technologies, which are required to bridge the gap in performance with the previous generations of wireless systems \cite{Boccardi, Andrews}. Cellular network densification is one such key solution that has been explored in 5G NR for achieving the required network capacity and spectral efficiency (SE) \cite{Bennis}. Ultra-dense network (UDN) is the culmination of network densification, where the number of base stations (BSs) approximates the number of users in both the spatial scale and magnitude \cite{Jafari,Sunliu}.

{The introduction of UDN has led to various unprecedented changes in the network performance.
	For instance, in the conventional sparse networks, the antenna heights of BSs and users have negligible impact on the channel quality 
	as the antenna height difference is much smaller than the link distance between a BS and its associated users. 
	However, with the UDN, the cell coverage area becomes significantly small, resulting in the small link distance. Therefore, the impact of BS and user antenna heights on the overall link distance (i.e., pathloss), channel fading distribution, the existence of line-of-sight (LoS) link in the BS-user channel etc. can no longer be ignored in the UDN scenario \cite{Atzeni}. Furthermore, due to the smaller pathloss in UDN, the changes in the channel fading distribution due to the existence or absence of the LoS link is also significant \cite{Cholee}. Hence, in UDN, we need to re-evaluate the strategies, used for the performance analysis and evaluation of the networks.}

{Due to above-mentioned features in UDNs, the performance analysis becomes more challenging and it also brings various new issues 
	in the network resource managements \cite{Hamouda, Tengliu, Shengliu}. 
	For instance, with increasing BS density, the rate of frequency reuse increases, which in turn increases the inter-cell interference (ICI) in UDN,
	and the rate of handover also increases due to the rapid changes in the coverage area of BSs \cite{Hasan}.
	Furthermore, the conventional received power/distance-based cell association strategies may not be efficient in UDN as there can be many BSs that provide the similar received signal power to a user.
}
{In addition, in reality, the UDN is likely to have the multi-tier structure, which is an inevitable consequence as BSs are densely deployed for the smooth integration of BSs into the existing cellular networks \cite{Hamouda, Tengliu}. 
	In such network architecture, BSs in different tiers would have different antenna heights, cell coverage area and transmission power.
	Consequently, an efficient interference management becomes more important, compared to the single-tier UDN \cite{Andrewshetnet,Dhillon, Kimlee}. It is more desirable especially for the users associated with lower tier BSs (i.e., BSs with smaller transmission power) since they can experience sever cross-tier interference from the higher tier BSs (i.e., BSs with higher transmission power).
	Therefore, the multi-tier UDN needs to be carefully designed for achieving better performance. 
	One solution for enhancing the higher performance in the multi-tier UDN is the coordinated multi-point (CoMP) transmission \cite{Sunliu, Mengzhao, menglu}. }

{The CoMP transmission is realized by the cooperation among multiple BSs to transmit the information to users. 
	Such transmission includes the dynamic point selection (DPS) \cite{3GPPr11}, the coordinated scheduling/ beamforming (CS/CB) \cite{Ngkim}, and the joint transmission (JT) \cite{Irmer}.  Among these techniques, the joint transmission is the most effective in terms of interference management, 
	but it gives more loads at each BS since multiple BSs need to transmit the same information for a user \cite{menglu}.
	However, in UDN, there are large number of BSs, even can be larger than the number of users \cite{Shengliu},
	so the joint transmission is expected to be used without overloadig BSs. 
	Hence, in this paper, we explore \emph{the possibility of implementing CoMP JT in multi-tier UDN}.}

{The CoMP JT strategies existing in literature have been developed, mainly for the conventional sparse cellular networks, 
	and implementing them directly in UDN creates several challenges. 
	For instance, the popular coherent JT strategy that utilizes the `$N$'-strongest/nearest BSs for CoMP transmission \cite{Nigam, Sunliu, RyuJlee}
	can have strong interfering BSs around a user. This is due to the fact that in UDN, the $(N+1)$-th strongest/nearest BS can still yield large interference since BSs are densely deployed. An alternative to this strategy is to determine cooperating BSs depending on various system parameters, e.g., fixed average received link power (ARLP) threshold \cite{Niejin}, the maximum per-link outage capacity \cite{GarciaShi, Zhouliu}, and the relative received link power (RRLP) threshold \cite{Fenghaenggi, Giovanidis}. The ARLP-based CoMP scheme determines the cooperating BSs as the ones with greater ARLP than a certain threshold\cite{Niejin}. However, this scheme can make a user have no cooperating BS, especially when the ARLP threshold is high. On the other hand, the maximum per-link outage capacity based CoMP scheme requires significant optimization overhead due to the presence of large number of BSs with similar ARLPs\cite{GarciaShi, Zhouliu}. The RRLP-based CoMP scheme determines the cooperating BSs as the ones with greater ARLP than a certain ratio of the strongest/nearest BS's ARLP  \cite{Liuzhou}. As a combined approach with ARLP based scheme, after determining two and three nearest BSs, the cooperating BSs are determined by their relative received power ratio in \cite{Fenghaenggi, Giovanidis}. However, \emph{the most works in CoMP JT did not consider the unique features of multi-tier UDN} such as the antenna heights of BSs and users and the random existence of LoS links, which are crucial in analyzing the performance in UDN.}
{Therefore, in this work, we consider the CoMP JT for a $K$-tier open-access UDN,
	where BSs of each tier use different transmission power and have different antenna heights to other tiers. 
	The BSs as well as the users are distributed by \acp{PPP}, densely over the network.
	We use the RRLP-based CoMP, which selects the cooperating BSs as the ones with greater ARLP than a certain ratio of ARLP of the strongest BS. 
	After deriving the coverage probability, we have also analyzed the area spectral efficiency (ASE), i.e., the spectral efficiency of users per unit area of the network, and the network energy efficiency (NEE), i.e., the ratio of ASE and the average network energy consumption per unit area. 
	{From the simulations, we have shown the effects of the BS densities and the number of cooperating BSs on the various performance metrics.}  
	To the best knowledge of us, this is the first work that analyzes the performance of CoMP JT by considering new features of UDN. 
	The contribution of this work can be summarized as follows. }

{
	\bi
	\item We newly develop an analytical framework for the RRLP-based CoMP scheme in a $K$-tier UDN
	by considering the new features of UDN. 
	Specifically, the probability of having LoS links, which is determined by the antenna height difference as well as the link distance, is considered, and the different channel models for LoS and NLoS links are used.
	\item By introducing the Gamma approximation for the received signal power from cooperating BSs, 
	we derive the coverage probabilities for general environment and for the special case with all NLoS links. 
	Although the derived coverage probability becomes complicated due to {the CoMP transmission in UDN}, 
	after careful examination, we have shown two different effects of the total BS density on the coverage probability. 
	\item We have revised the conventional ASE to make it more suitable for CoMP JT scenarios in UDN. 
	First, we define the transmit ASE (Tx ASE) by counting the links from cooperating BSs (i.e., multiple transmitters) as one valid link in the definition of ASE since those BSs actually transmit the same information. 
	We also introduce the receiver ASE (Rx ASE) by considering the SE per receiver (i.e., user) instead of the SE per transmitter (i.e., BS)
	since this can be more valid when the number of BSs is larger than that of users like UDN environment.
	\item Using the ASEs, we define and compute the NEE for a multi-tier UDN by considering not only the transmission power but also various transceiver circuit power consumptions (which is often ignore in existing works despite of non-negligible values of consumption). 
	After investigating the NEE through simulations, the useful insights on the energy-efficient CoMP transmission design are provided, which also further motivates the use of the RRLP-based CoMP, compared to the case without CoMP.
	\ei
}

The remainder of this paper is organized as follows: Section~\ref{sec:models} introduces the system model, the RRLP based CoMP transmission design, as well as the performance metrics. In Section~\ref{sec:perfanalysis}, we present the detailed coverage probability analysis by deriving the approximate distribution of the received signal power. In Section~\ref{sec:results}, we discuss the simulation results. Conclusions are presented in Section~\ref{sec:conclusion}.

\begin{table}
	\caption{Notations used throughout the paper.} \label{table:notation}
	\vspace{-0.25cm}
	\begin{center}
		\rowcolors{2}
		{cyan!15!}{}
		\renewcommand{\arraystretch}{1.0}
		\begin{tabular}{| c | p{6.5cm} || c | p{6.5cm} |}
			\hline 
			{\bf Notation} & {\hspace{2.5cm}}{\bf Definition} & {\bf Notation} & {\hspace{2.5cm}}{\bf Definition}
			\\
			\midrule
			\hline
			$\bm \Phi_{b,j}$ & PPP for BS distribution in the $j$-th tier & $\lambda_{b,j}$ & Spatial Density of BSs in the $j$-th tier \\ 
			$\bm \Phi_{u}$ & PPP for user distribution in the network & $\lambda_{u}$ & Spatial Density of users in the network \\ 
			$h_{b,j}$ & Antenna height of BSs in the $j$-th tier & $h_{u}$ & Antenna height of users in the network \\ 
			$\nc \in \{ \tL, \tN \}$ & Indicator of LoS/NLoS channel & $\ack$ & Pathloss exponent for channel $\nc$ \\ 
			$\nco \in \{ \tL, \tN \}$ & {Indicator of LoS/NLoS channel for the main link} & $x_{k,o}$ & {Link distance from the main link BS to the typical user} \\ 
			$g_{j,i}^{(\nc)}$ & {Fading gain between the typical user and BS $i$ with channel $\nc$ in the $j$-th tier} & $x_{j,i}$ & {Link distance from BS $i$ in the $j$-th tier to the typical user} \\ 
			$\varsigma_{j}$ & Average downlink transmit Power of BSs in the $j$-th tier & $y_{j,i}$ & {Horizontal distance between the typical user and BS $i$ in the $j$-th tier} \\ 
			$m^{(\nc)}$ & Shape parameter of Nakagami-m fading distribution for channel $\nc$ & $p^{\ncb}(x)$ & Probability of having channel $\nc$ between a BS and user at link distance $x$ \\ 
			$\varepsilon$ & Fraction of network area covered by buildings/blockage & $\rho$ & Average building/blockage height \\ 
			$\Upsilon$ & Average number of buildings per unit area & $C_0$ & Coherence block duration \\ 
			$B$ & Communication bandwidth & $\frac{1}{\kappa_j}$ & Power amplifier efficiency of a BS in the $j$-th tier \\ 
			$L_{b,j}$ & Computation efficiency of a BS in the $j$-th tier & $\bar{U}_{j}$ & Average number of associated user with a BS in the $j$-th tier \\ 
			$P_{\text{rf},j}^{(b)}$ & Antenna power consumption at a BS in $j$-th tier & $P_{\text{fix},j}$ & Fixed power consumption at a BS in the $j$-th tier \\ 
			$P_{\text{rf}}^{(u)}$ & Antenna power consumption at a user & $P_{\text{rate}}$ & Rate dependent power consumption at user \\ 
			$n_{j}^{(\nc)}$ & Number of BSs in the $j$-th tier associated with the typical user with channel $\nc$ & $\gamma_{\nco,k}$ & Received SIR at the typical user having a main link BS with channel $\nco$ from the $k$-th tier \\ 
			$\mathcal{A}_{k}^{\ncob}$ & Probability that the main link BS with channel $\nco$ from the $k$-th tier & $f_{X}(.)$ & Probability density distribution of random variable $X$ \\ 
			\hline 
		\end{tabular}
	\end{center}\vspace{-0.63cm}
\end{table}%

\vspace{-0.6 cm}

\section{System Model}\label{sec:models}
In this section, we first present the system model for a open-access multi-tier UDN, including the model for LoS channel probability and the CoMP transmission strategy. We then present the performance metrics considered in this paper.

\vspace{-0.4 cm}

\subsection{UDN Network and Channel Model}
We consider a $K$- tier UDN, where the base stations (BSs) from different tiers are distributed according to independent homogeneous Poisson point process (HPPP), and are distinguished based on their individual downlink transmission power as well as the BS antenna heights. For instance, the distribution of BSs in the $j$-th tier follows an independent HPPP $\bm \Phi_{b,j}$, with the spatial density $\lambda_{b,j}$, and have BS antenna height $h_{b,j}$ and downlink transmit power $\varsigma_j$ ($j = 1, 2, \ldots, K$). We also assume that the distribution of users in the network follows an independent HPPP $\bm \Phi_u$ with the spatial density $\lambda_u$. 
\par The channel between the BS and the typical user can be a LoS channel, as long as the line-of-sight path between them is not obstructed by a blockage (e.g. buildings in dense urban environment). However, the existence of such LoS path in the channel cannot be guaranteed due to the random nature of user locations in the network (e.g., due to the mobility of users). Hence, depending on the locations and heights of the BSs and buildings in the network, we characterize the probability of existence of LoS path in the channel. %
\par Several models have been adopted to characterize this LoS probability as a function of link distance, in a simple mathematical form for analysis \cite{3GPPr112, lardner, IMT2020propagate}. However, most of these models do not consider the antenna height of BS and user, which give significant impact on the BS-user distance in UDN scenarios. Recently, in \cite{Kimlee}, a more complete version of the LoS probability model has been presented using the results in \cite{Vazebai} and the ITU blockage model. {Using this model in our scenario, the probability of having a LoS/NLoS link channel between BS $i$ in the $j$-th tier and the typical user is given by}
\vspace{-0.3 cm}
{\begin{align}
\label{eq:losfunc}
p^{\ncb}(x_{j,i}) & = \left\{ \begin{array}{ll} \left( 1 - \sqrt{\frac{\pi}{2}}\frac{\rho}{h_j} \left[ \erf\left( \frac{h_{b,j}}{\rho\sqrt{2}}\right) - \erf\left( \frac{h_{u}}{\rho\sqrt{2}} \right) \right] \right)^{\sqrt{\varepsilon \, \Upsilon \{x_{j,i}^2 - (h_{b,j} - h_{u})^2\} }} \, , & \text{if }\nc = \tL \\
1 - \left( 1 - \sqrt{\frac{\pi}{2}}\frac{\rho}{h_j} \left[ \erf\left( \frac{h_{b,j}}{\rho\sqrt{2}}\right) - \erf\left( \frac{h_{u}}{\rho\sqrt{2}} \right) \right] \right)^{\sqrt{\varepsilon \, \Upsilon \{x_{j,i}^2 - (h_{b,j} - h_{u})^2\} }} \, , & \text{if } \nc = \tN
\end{array} \right.
\end{align}}
\noindent where $\nc \in \{ \tL, \tN \}$ denotes the channel environment, i.e.,  $\nc = \tL$ for LoS link, and $\nc = \tN$ for NLoS link, and $\erf(z) = \frac{1}{\sqrt{\pi}} \, \int_{-z}^{z} \, e^{-t^2} \, dt$. From \eqref{eq:losfunc}, it is evident that the LoS link probability (i.e., $\nc = \tL$), $p_{\tL}(x)$, decreases with the link distance $x$, since higher value of $x$ increases the probability of inclusion of a blockage/ building of sufficient height to block the LoS path in the channel. The probability of NLoS channel (i.e., $\nc = \tN$) is similarly given by $p_{\tN}(x) \Define 1 - p_{\tL}(x)$.
{The power of the signal received from BS $i$ in $j$-th tier at the typical user is therefore given by}
\vspace{-0.3 cm}
\begin{align}
\label{eq:rxpower}
P^{\ncb}_j(x_{j,i}) & \Define \varsigma_{j} \, \, g^{\ncb}_{j,i} x_{j,i}^{-\ack} \, ,
\end{align}
\noindent where $g^{\ncb}_{j,i}$ and $\ack \geq 2$ respectively denote the small scale fading gain and the pathloss exponent of the channel $\nc$, and $x_{j,i}$ denotes the corresponding link distance of this BS from the typical user. Here, $x_{j,i}$ is given by $x_{j,i} \Define \sqrt{y_{j,i}^2 + h_j^2}$, where $y_{j,i}$ is the horizontal distance between the user and the BS, and $h_j \Define h_{b,j} - h_u$ is the difference in antenna heights between the $j$-th tier BSs and the typical user. Without loss in generality, we assume that the small scale fading gain in a channel $\nc$ follows the Nakagami-\textit{m} distribution with parameter $m^{\ncb}$ and mean value $1$, i.e., $g^{\ncb}_{j,i} \sim \Gamma \left(m^{\ncb}, \frac{1}{m^{\ncb}} \right)$ \cite{Atzeni, Cholee}.\footnote[1]{Note that the generic Nakagami-\textit{m} distribution can be used to characterize any generalized form of fading gain in the wireless channel. For instance, in a conventional NLoS channel (i.e., $\nc = \tN$), the fading is assumed to be Rayleigh distributed (i.e., $\mN = 1$). In general, we have, $\mL \geq \mN$.}
\vspace{-0.5 cm}

\subsection{Relative Received Link Power Based CoMP Transmission}\label{rrpcomp}
We consider a relative received link power (RRLP) based approach for determining the set of associated/ cooperating BSs for the typical user 
{under the ideal backhaul link environment.\footnote[2]{{In this work, we consider the ideal backhaul link environment (i.e., with infinite capacity) to focus on the performance analysis of the access link (between BSs and the user), and leave the analysis on the effect of the limited backhaul link capacity in UDN as a future work.}}}
{In this approach, we first check the average received link power (ARLP) from all the neighbouring BSs in different tiers, and compare them to determine the BS with the strongest ARLP. We denote this BS as the BS \lq$o$' in the cooperation set and term the link between this BS and the typical user as the \emph{main link}. We therefore denote the main link distance as $x_{k,o}$ and its corresponding channel environment as $\nco$ ($\nco \in \{ \tL, \tN \}$). In order to determine whether a neighbouring BS is a cooperating BS for the typical user, we compare its ARLP to that of the main link BS, and check whether the ratio of their ARLPs exceeds a pre-defined threshold value.\footnote[3]{Note that in general any BS in the network can be considered a neighbouring BS to the typical user. However, such assumption becomes unnecessary in case of large networks, where BSs at sufficiently far away from the user have negligible ARLP, and therefore would have little impact in selecting the main link BS. Therefore, for practical implementation, we can limit the number of neighbouring BSs by imposing a suitable constraint on either the ARLP or on the horizontal distance between the typical user and the BSs. Such restriction however does not impact the cooperating BS set selection, as long as this threshold remains sufficiently small.} For instance, when the typical user has its main link BS in the $k$-th tier, the neighbouring BS $i$ in the $j$-th tier with channel $\nc$ to the typical user participates in the cooperation set, if it satisfies the following condition:}
\vspace{-0.0 cm}
{\begin{align}
\label{eq:compcondjk}
\frac{\bar{P}^{\ncb}_j (x_{j,i})}{\bar{P}^{\ncob}_k (x_{k,o})} & \geq \eta_{j,k} \, ,
\end{align}}
\noindent {where $\bar{P}^{\ncb}_j (x_{j,i}) = \Eb{P^{\ncb}_j(x_{j,i})} = \varsigma_j \, x_{j,i}^{-\ack}$ denotes the ARLP of the BS $i$ in $j$-th tier, and $\eta_{j,k}$ ($j = 1, 2, \ldots, K$) denotes the relative link power ratio (RLPT) for the $j$-th tier BSs, given that the main link BS is from the $k$-th tier. }

Note that $\eta_{j,k}$ denotes the \emph{minimum fraction of the main link ARLP}, required by any BS in the $j$-th tier to be included in the cooperation set. Clearly, if $\eta_{j,k}$ is set at a higher value, the probability of inclusion of a $j$-th tier BS in the cooperation set automatically decreases. The same also holds when the main link ARLP increases for a given $\eta_{j,k}$. In other words, using $\eta_{j,k}$ as a threshold, we can dynamically adjust the number of cooperating BSs in each tier of the multi-tier UDN for a user (see Remark~\ref{remarkNbs}). Note that such a transmission strategy not only reduces the number of interfering BSs with strong ARLP, but also controls the number of cooperating BSs. In other words, any BS, which does not have sufficiently strong ARLP, is not included in the cooperation set, since it only provides incremental change in the overall system performance. Hence, this strategy prevents any non-essential signal strength increase at the user in determination of its cooperating BS set, thus preventing any excess energy consumption in the network.\footnote[4]{{It can be shown that for average number of cooperating BSs fixed at $3$, there is $95\%$ probability that the maximum number of cooperating BSs would be less than or equal to $7$. This shows that the probability that the CoMP set size is very large with the RRLP-based CoMP transmission is negligibly small.}}
\subsubsection{SIR for Downlink Transmission} \label{sirsection}
{In order to analyze the performance of the RRLP-based  CoMP strategy, we consider the coherent joint transmission scenario in the downlink. In other words, the BSs, associated with the typical user, cooperate to transmit information symbols to the user in a synchronized manner. We also assume that the channel state information of the associated users are known at the BSs, and consider the conjugate beamforming based precoding, as in \cite{Nigam}. Thus, when the main link BS to the typical user is in the $k$-th tier with channel $\nco$ and link distance $x_{k,o}$, the overall received signal at the typical user in the interference limited environment is given by\footnote[5]{Note that inclusion of noise in the system model would further complicate the performance analysis, while the trend observed for the overall performance would be similar \cite{Nigam, Dhillon}. Therefore, we have limited our analysis to the interference limited environment only.}}
\vspace{-0.1 cm}
{
	{\begin{align}
			\label{eq:rxsigcb}
			\nonumber \mathcal{Z}_o 
			& = \sum\limits_{j=1}^{K} \sum\limits_{\nc \in \{ \tL, \tN \} } 
			\sum\limits_{i \in \bm{\phi}_{b,j}^{\ncb} } 
			\underbrace{\sqrt{\varsigma_{j}} \, \, q^{\ncb}_{j,i} x_{j,i}^{-\ack/2} w_{j,i}^{\ncb} u_o}_{\substack{\text{signal from the associated BSs}\\ \text{in the $j$-th tier}}} \, \1{x_{j,i} \leq \theta_j^{\ncb}(x_{k,o},\nco) \eta_{j,k}^{-\frac{1}{\ack}}} \,  \\
			& +
			%
			%
			\underbrace{ \sqrt{\varsigma_{j}} \, \, q^{\ncb}_{j,i} x_{j,i}^{-\ack/2} w_{j,i}^{\ncb} u_{j,i} }_{\text{signal from interfering BSs}}\, \1{x_{j,i} > \theta_j^{\ncb}(x_{k,o},\nco) \eta_{j,k}^{-\frac{1}{\ack}}} \, 
\end{align}}}
\noindent {{where $\bm{\phi}_{b,j}^{\ncb}$ is the index set of BSs in the $j$-th tier with channel $\nc$ to the typical user,}
	$u_o$ is the transmitted symbol of the main link BS, $u_{j,i}$ is the independent transmitted symbol from an interfering BS,\footnote[6]{Note that without loss of generality, we have assumed the transmitted symbol power to be $1$ and all channels between the BSs and typical user are independent.} $q_{j,i}^{\ncb}$ is the channel fading gain with the typical user for the BS $i$ in $j$-th tier, and $w_{j,i}^{\ncb} = \frac{(q_{j,i}^{\ncb})^{\ast}}{\left| q_{j,i}^{\ncb} \right|}$ denotes its corresponding normalized precoder. 
	{Here 
		$\theta_j^{\ncb}(x_{k,o},\nco) \Define \left(\frac{\varsigma_j }{\varsigma_k} \right)^{\frac{1}{\ack}} \, x_{k,o}^{\frac{ \aco }{ \ack }}$, and $\1{.}$ denotes the mathematical indicator function.} and $g_{j,i}^{\ncb} = |q^{\ncb}_{j,i} w_{j,i}^{\ncb}|^2 = |q^{\ncb}_{j,i}|^2$. Clearly, the total received signal power at the typical user can be written as  }
\vspace{-0.2 cm}
{\begin{align}
\label{eq:sigdef}
P_{\text{comp}}^{\ncob}(x_{k,o}) & = \left | \underbrace{\sqrt{P_k^{\ncob}(x_{k,o})}}_{\text{signal from main link BS}} \hspace{-0.2 cm} + \, \sum\limits_{j=1}^{K} \sum\limits_{\nc \in \{ \tL, \tN\}} \sum\limits_{i \in \bm \phi_{b,j}^{(\nc)}} \underbrace{\sqrt{P_j^{\ncb}(x_{j,i}) } \1{x_{j,i} \leq \theta_j^{\ncb}(x_{k,o},\nco) \eta_{j,k}^{-\frac{1}{\ack}}}}_{\text{signal from other associated BSs in the $j$-th tier}} \right|^2 \, ,
\end{align}}
\noindent {We further note that any BS outside the cooperation set of the typical user acts as an interferer to that user, and therefore the overall interference power received at the user is given by}
\vspace{-0.25 cm}
{\begin{align}
\label{eq:intfdef} 
I^{\ncob}(x_{k,o}) & = \sum\limits_{j=1}^{K} \sum\limits_{\nc \in \{ \tL, \tN\}} \sum\limits_{i \in \bm \phi_{b,j}^{(\nc)}} P_j^{\ncb}(x_{j,i}) \1{x_{j,i} > \theta_j^{\ncb}(x_{k,o},\nco) \eta_{j,k}^{-\frac{1}{\ack}} } \, . 
\end{align}}
{Then, the overall received signal-to-interference power ratio (SIR) in an interference-limited environment is given by}
\vspace{-0.4 cm}
{\begin{align}
\label{eq:sirdef}
\gamma_{k}^{\ncob}(x_{k,o}) &  = \frac{P_{\text{comp}}^{\ncob}(x_{k,o})}{I^{\ncob}(x_{k,o})} \, .
\end{align}}
\vspace{-0.7 cm}
\subsubsection{Main Link Distance Distribution}
In section~\ref{rrpcomp} we show that the performance of the RRLP based CoMP transmission strategy in the multi-tier UDN scenario is heavily dependent on the tier and nature of channel (LoS/ NLoS) of the main link BS. Therefore, prior to introducing the performance metrics under consideration, we first present an expression for the probability density function (pdf) of the main link distance in this CoMP transmission scenario. %
%
%
\begin{lemma}\label{mainlinkpdfc}
	\normalfont For the main link BS in $k$-th tier, let us denote the main link distance as $X_{k,o}^{\ncob}$. The pdf of $X_{k,o}^{\ncob}$ is given by
	\vspace{-0.6 cm}
	\begin{align}
		\label{eq:mainlinkpdf}
		f_{X_{k,o}^{\ncob}}(r) & = \frac{2 \pi \lambda_{b,k}}{\mathcal{A}_{k}^{\ncob}} \, r \, p^{\ncob}(r) \, e^{ -2\pi \,\sum\limits_{j=1}^{K} \lambda_{b,j}  \sum\limits_{\nc \in \{ \tL, \tN \} }  \int_{h_j}^{\theta_j^{\ncb}(r,\nco)} t \, p^{\ncb}(t) \,  dt} \, .
	\end{align}
	\noindent where $p^{\ncb}(x)$ is the probability of having channel environment $\nc$ for a link distance $x$, and $\theta_j^{\ncb}(r,\nco) = \left(\frac{\varsigma_j }{\varsigma_k} \right)^{\frac{1}{\ack}} \, r^{\frac{ \aco }{ \ack }}$. Here $\mathcal{A}_{k}^{\ncob}$ is the tier association probability for the main link BS, and is given by
\vspace{-0.5 cm}
\begin{align}
	\label{eq:mainlinktypeprob}
	\mathcal{A}_{k}^{\ncob} & = \int_{h_k}^{\infty} 2 \pi \lambda_{b,k} \, x \, p^{\ncob}(x) \,  e^{ - 2\pi \sum\limits_{j=1}^{K} \lambda_{b,j} \sum\limits_{\nc \in \{ \tL, \tN \} } \int_{h_j}^{\theta_j^{\ncb}(x,\nco)} t \, p^{\ncb} (t) dt } \, dx .
\end{align}
\end{lemma}
\begin{IEEEproof}
	See Appendix~\ref{app:pdfcproof}.
\end{IEEEproof}

\subsubsection{Number of Cooperating BSs}

The number of cooperating BSs of the typical user in each tier of the network depends both on the main link BS (i.e., its ARLP and its tier), and also on the RLPT corresponding to that tier. For instance, when the typical user has a main link BS in the $k$-th tier, the number of cooperating BSs in the $j$-th tier is
\vspace{-0.35 cm}
\begin{align}
\label{eq:compsetsizej}
N_j^{\ncob}(x_{k,o}) & =  \sum\limits_{\nc \in \{ \tL, \tN \} } \sum\limits_{i \in \bm \phi_{b, j}^{\ncb} } \1{x_{j,i} \leq   \theta_{j}^{\ncb}(x_{k,o},\nco) \, \eta_{j,k}^{-\frac{1}{\ack}} \,       } \, .
\end{align}
\noindent Thus, the total number of cooperating BSs for the typical user in the network is given by 
\vspace{-0.35 cm}
\begin{align}
\label{eq:compsetsize}
N^{\ncob}(x_{k,o}) & = \sum\limits_{j = 1}^{K} N_j^{\ncob}(x_{k,o}) \, .
\end{align}
\vspace{-0.5 cm}
\par
\begin{figure}[t!]
\vspace{-1.4 cm}
    \begin{center}   
    { 
	 \includegraphics[width=0.5\columnwidth]{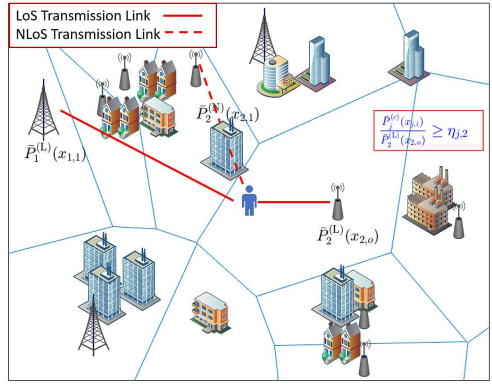} 
    }
    \end{center}
    \caption{
    		{An example scenario of the RRLP based CoMP transmission strategy in a 2-tier (macro and micro) UDN.} 
		 }
   \label{fig:NetworkModel} \vspace{-0.5 cm}
\end{figure}

 For instance, in Fig.~\ref{fig:NetworkModel}, we show the associated BSs cooperating to serve the typical user. It is observed that this user is served by two BSs with LoS channel to the user (ARLPs are $\bar{P}_2^{(\tL)}(x_{2,o})$ and $\bar{P}_1^{(\tL)}(x_{1,1})$ respectively), and one BS with NLoS channel (ARLP is $\bar{P}_2^{(\tN)}(x_{2,1})$). Clearly, the main link BS for this user is in 2nd tier and has a LoS channel to the user. Thus, we have $N_1^{(\tL)}(x_{2,o}) = 1$, $N_2^{(\tL)}(x_{2,o}) = 2$, and $N^{(\tL)}(x_{2,o}) = N_1^{(\tL)}(x_{2,o}) + N_2^{(\tL)}(x_{2,o}) = 3$.
\begin{lemma}
\label{avgbssetsize}
\normalfont The average number of cooperating BSs for the typical user with the RRLP-based CoMP transmission strategy is given by
\vspace{-0.3 cm}
\begin{align}
\label{eq:avgbssize}
N_{\text{avg}} & = \sum\limits_{k=1}^{K} \sum\limits_{\nco \in \{ \tL, \tN \} } \, \mathcal{A}_{k}^{\ncob} \int_{h_k}^{\infty}  f_{X_{k,o}^{\ncob}}(r) \, \sum\limits_{j=1}^{K}  2\pi \lambda_{b,j} \sum\limits_{\nc \in \{ \tL, \tN \} } \int_{h_j}^{\theta_j^{\ncb}(r,\nco) \eta_{j,k}^{-\frac{1}{\ack}}} t \, p^{\ncb}(t) \, dt \,dr ,
\end{align}
\noindent  where, $\mathcal{A}_{k}^{\ncob}$ is defined in \eqref{eq:mainlinktypeprob}.
\end{lemma}
\begin{IEEEproof}
See Appendix~\ref{app:avgbssetsizep}. 
\end{IEEEproof}
\begin{remark}
	\label{remarkNbs}\normalfont 
	{\emph{(Relation between $\eta_{j,k}$ and $\lambda_b$)} From \eqref{eq:avgbssize}, it is clear that the BS densities and the RLPT parameters, $\eta_{j,k}$ are functions of each other, but it is difficult to obtain their explicit relationship. However, for a simplified scenario (like when $K = 1$ (i.e., single tier), and $p^{(\tL)}(x) = 0$, for all $x$), we can obtain the relation between $\eta_{1,1}$, and $\lambda_{b,1}$ as $\lambda_{b,1} =  \frac{N_{\text{avg}} \eta_{1,1}^{\frac{2}{\aN} } - 1}{\pi h_{b,1}^2 \left(1 - \eta_{1,1}^{\frac{2}{\aN} }\right) }$. From this expression, it can be easily shown that $\frac{\partial \eta_{1,1}}{\partial \lambda_{b,1}} > 0$, for all $N_{\text{avg}} >1$. This means for a fixed average number of cooperating BSs, $\eta_{1,1}$ monotonically increases with $\lambda_{b,1}$, and vice versa. This shows that as the total BS density increases, the relative received power threshold also needs to increase, in order to keep the total average number of cooperating BSs fixed. This conclusion is also verified for the multi-tier UDN as shown in Table.~\ref{table:etalim}.}
\end{remark}

\vspace{-0.5 cm}
\subsection{Performance Metrics}\label{metrics}
\vspace{-0.4 cm}
In order to analyze the performance of the RRLP based CoMP transmission strategy, we define the following metrics: (a) coverage probability, (b) area spectral efficiency (ASE), and (c) network energy efficiency (NEE). The coverage probability metric determines the probability of achieving a desired received SIR threshold. The ASE metric gives a measure of the total network throughput per unit area, while the NEE determines the information rate per unit energy consumption in the network.
\vspace{-0.2 cm}
\subsubsection{Coverage Probability} \label{covpdef}
From our discussion in section~\ref{rrpcomp}, it is clear that the received SIR at the typical user depends on the tier, the nature of channel as well as the link distance of its main link BS. Therefore, for a given desired SIR threshold $\widetilde{\gamma}$, the overall downlink coverage probability of the typical user for RRLP-based CoMP transmission in a $K$-tier open access UDN  is given by
\vspace{-0.5 cm}
\begin{align}
\label{eq:covpdef}
\mathcal{P}_{\text{cov}}(\widetilde{\gamma}, \bm \lambda_b) & \Define  \sum\limits_{k=1}^{K} \sum\limits_{\nco \in \{\tL, \tN \}} \mathcal{A}_{k}^{\ncob} \int_{h_k}^{\infty} f_{X_{k,o}^{\ncob}}(r) \, \underbrace{\Pb{\gamma_{k}^{\ncob}(r) \geq \widetilde{\gamma}}}_{\Define \,\, p_{\text{cov}}(r,k,\nco,\widetilde{\gamma})} dr \, ,
\end{align}
\noindent where $\bm \lambda_b \Define (\lambda_{b,1}, \lambda_{b,2}, \cdots, \lambda_{b, K})^T$, and $\gamma_{k}^{\ncob}(r)$ denotes the received SIR of the typical user, having main link BS in the $k$-th tier, with link distance $r$, and channel $\nco$ (see \eqref{eq:sirdef}). 
{\begin{table}
		\caption{{RLPT parameter, $\eta_{j,k}$, with different total BS density, $\lambda_b$, and $N_{\text{avg}}$, for a 2-tier UDN, when $\frac{\lambda_{b,1}}{\lambda_{b}} = 0.2$, $\frac{\lambda_{b,2}}{\lambda_{b}} = 0.8$, and $\eta_{j,k} = \eta$, for all $j,k$.}	} \label{table:etalim}
		\vspace{-0.25cm}
		\begin{center}
			\rowcolors{2}
			{cyan!15!}{}
			\renewcommand{\arraystretch}{1.0}
			{\begin{tabular}{| c | c | c | c | c | c | c |}
					\hline 
					{$\lambda_b$} (/m\textsuperscript{2}) & $10^{-5}$ & $10^{-4}$ & $10^{-3}$ & $5 \times 10^{-3}$ & $10^{-2}$ & $5 \times 10^{-2}$
					\\
					\hline 
					$\eta$ (in dB) for $N_{\text{avg}} = 2$ & -7.70 & -5.85 & -4.56 & -1.74 & -1.02 & -0.22 \\
					$\eta$ (in dB) for $N_{\text{avg}} = 3$ & -12.22 & -9.20 & -7.96 & -3.19 & -1.92 & -0.43 \\
					\hline 
			\end{tabular}}
		\end{center}\vspace{-0.9cm}
\end{table}}%
\subsubsection{Area Spectral Efficiency (ASE)}\label{asedefall}
In the literature, the area spectral efficiency (ASE) metric has been widely used for network performance analysis, especially in the conventional single BS association scenario. The ASE measures the spectral efficiency (SE) of users per unit area of the network, and it is defined as the product of BS density ($\lambda_b$) and the average per-link SE, with a minimum required operational SIR, i.e., $\Eb{\log_2(1 + \text{SIR}) \1{\text{SIR} \geq \widetilde{\gamma}} }$ \cite{Ding,Ding2, Alammouri}. Utilizing the definition of average rate in \cite{Josang}, this definition has also been extended to the multi-tier open access network model \cite{Kimlee}.  

\par We however note that the above definition of ASE is valid only when the BS density is significantly smaller than the user density, since it is defined with the assumption of active transmission of all the BSs. We also note that this definition is not compatible with the scenario, where a user can associate with multiple BSs (e.g. CoMP). In CoMP transmission, the received signal power at the user is the power of sum of the received signals from all cooperating BSs (see \eqref{eq:sigdef}). In this scenario, one feasible method for defining the ASE ise to consider the per-user SE per unit area and define the ASE as a function of the user density. This definition although holds true in the high BS density scenario, it may not be accurate in the conventional wireless networks, where the BS density is significantly smaller compared to the user density. Therefore, in order to bridge this gap in performance measurement at varying BS densities, in this paper, we define ASE in both ways. The definition of ASE utilizing the BS density (i.e. the transmitter density in the downlink) is denoted as the \emph{transmit ASE (Tx ASE)}, and the definition of ASE utilizing the user density is termed as the \emph{receive ASE (Rx ASE)}.
\begin{definition}\label{txasedef}
\normalfont The \emph{transmit ASE (Tx ASE)} of downlink CoMP transmission for a $K$-tier open access network is defined as 
\vspace{-0.4 cm}
\begin{align}
\label{eq:tasedef}
\mathcal{S}_t(\widetilde{\gamma}, \bm \lambda_b) & \Define \sum\limits_{j=1}^{K} \lambda_{b,j} \mathcal{S}_j(\widetilde{\gamma}, \bm \lambda_b) \, .
\end{align}
\noindent where $\mathcal{S}_j(\widetilde{\gamma}, \bm \lambda_b)$ is the per-transmission link SE  provided by the $j$-th tier BSs. Note that this tier-wise per-transmission link SE also depends on the tier and channel of the main link BS in the RRLP based CoMP transmission strategy, so it can be written as 
\vspace{-0.2 cm}
\begin{align}
\label{eq:tasejtier}
\mathcal{S}_{j}(\widetilde{\gamma}, \bm \lambda_b) & = \sum\limits_{k=1}^{K} \sum\limits_{\nco \in \{ \tL, \tN \} } \mathcal{A}_{k}^{\ncob} \mathcal{R}_{j,k}^{\ncob}(\widetilde{\gamma}) \, ,
\end{align}
\noindent where $\mathcal{R}_{j,k}^{\ncob}(\widetilde{\gamma})$ denotes the average contribution of a cooperating BS from the $j$-th tier in the per-link SE of the typical user \cite{Fenghaenggi}, given that its main link BS is in the $k$-th tier, and has a channel $\nco$. Finally, in order to obtain an estimate of per-link SE for each tier of cooperating BSs, we define $\mathcal{R}_{j,k}^{\ncob}(\widetilde{\gamma})$ as 
\vspace{-0.25 cm}
\begin{align}
\label{eq:txase}
& \mathcal{R}_{j,k}^{\ncob}(\widetilde{\gamma})  \Define   \int_{h_k}^{\infty} \,  \E\left[ \underbrace{\frac{ \sum\limits_{\nc \in \{\tL, \tN \} }  \, n_j^{\ncb}}{N^{\ncob}(r)}}_{\Define \, \psi_{j,k}^{\ncob} }  \underbrace{\frac{\log_2(1 + \gamma_{k}^{\ncob}(r))}{N^{\ncob}(r)}}_{\text{per-link SE}} \1{\gamma_{k}^{\ncob}(r) \geq \widetilde{\gamma}} \right] f_{X_{k,o}^{\ncob}}(r) \, dr 
\end{align}
\noindent where $\psi_{j,k}^{\ncob}$ represents the fractional contribution of the $j$-th tier BSs in $\mathcal{R}_{j,k}^{\ncob}(\widetilde{\gamma})$. Thus, the total Tx ASE for the CoMP transmission scenario can be given by
\begin{align}
\label{eq:txaseredef}
\mathcal{S}_{t}(\widetilde{\gamma}, \bm \lambda_b) \, = \sum\limits_{j=1}^{K} \lambda_{b,j} \sum\limits_{k=1}^{K} \sum\limits_{\nco \in \{ \tL, \tN \} } \mathcal{A}_{k}^{\ncob} \mathcal{R}_{j,k}^{\ncob}(\widetilde{\gamma}) \, .
\end{align}
Note that this definition of Tx ASE converges to the ASE definitions in \cite{Ding, Ding2, Alammouri, Kimlee} for the single BS association scenario (i.e., when $N^{\ncob}(r) =1$).
\end{definition}
\begin{definition}\label{rxasedef}
\normalfont The \emph{receive ASE (Rx ASE)} of the downlink $K$-tier open-access CoMP network is given by
\vspace{-0.4 cm}
\begin{align}
\label{eq:rasedef}
\mathcal{S}_r(\widetilde{\gamma}, \bm \lambda_b) & \Define \lambda_u \mathcal{R}_u(\widetilde{\gamma}, \bm \lambda_b) 
\end{align}
\noindent where $\mathcal{R}_u(\widetilde{\gamma}, \bm \lambda_b)$ is the per-user SE, which is defined as 
\begin{align}
\label{eq:puse}
\mathcal{R}_u(\widetilde{\gamma}, \bm \lambda_b) & = \sum\limits_{k=1}^{K} \sum\limits_{\nco \in \{ \tL, \tN \} } \mathcal{A}_{k}^{\ncob} \int_{h_k}^{\infty} \, f_{X_{k,o}^{\ncob}}(r) \, \Eb{\log_2(1 + \gamma_{k}^{\ncob}(r)) \1{\gamma_{k}^{\ncob}(r) \geq \widetilde{\gamma}} } dr \, .
\end{align}
\end{definition}

\subsubsection{Network Energy Efficiency}\label{needef}
The \emph{network energy efficiency (NEE)} metric can be defined as the ratio of the ASE and the total average network energy consumption (NEC) per unit area of the network. The NEC in turn depends on both the transmission energy consumption as well as the circuit energy consumption at the BSs and users. In the following, we first characterize the transmission and circuit power consumption parameters at the BSs and users, and then define the NEE. For the downlink transmission, the total average power consumption at the $j$-th tier BS is given by \cite{HoydisEmil}
\vspace{-0.4 cm}
\begin{align}
\label{eq:bspcp}
P_{\text{BS}, j} & = P_{\text{rf},j}^{(b)} \, + \, P_{\text{fix},j} \, + \, \frac{1}{\kappa_j}\varsigma_j \, + \, \frac{3 \, B}{C_0 \, L_{b,j}} \bar{U}_j \, ,
\end{align}
\noindent where $P_{\text{rf},j}^{(b)}$ is the antenna power consumption, $P_{\text{fix},j}$ is the fixed power consumption (due to control signalling, backhaul, site cooling etc.), and $\frac{1}{\kappa_j}$ means the efficiency of the power amplifier (PA) at the $j$-th tier BS. The last term in \eqref{eq:bspcp} is the signal processing power consumption at the BS (due to matched filter processing) \cite{HoydisEmil}, which depends on the total bandwidth ($B$), the duration of coherence block (in symbols), $C_0$, the computational efficiency of the BS ($L_{b,j}$), and the average number of users associated with a BS in the $j$-th tier, $\bar{U}_j$. Similarly, the total power consumption at any user can be characterized as 
\vspace{-0.4 cm}
\begin{align}
\label{eq:uepcp}
P_{\text{UE}} = P_{\text{rf}}^{(u)} \, + \, \mathcal{R}_u(\widetilde{\gamma}, \bm \lambda_b) \, B \, P_{\text{rate}} \, ,
\end{align}
\noindent where $P_{\text{rf}}^{(u)}$ is the antenna power consumption at the user and $P_{\text{rate}}$ is the rate dependent power consumption. Thus, the total average power consumption per unit area is given by
\vspace{-0.3 cm}
\begin{align}
\label{eq:pneca}
P_{\text{nec}} (\widetilde{\gamma}, \bm \lambda_b) & = \sum\limits_{j=1}^{K} \lambda_{b, j} \, P_{\text{BS},j} \, + \, \lambda_u \, P_{\text{UE}} \, .
\end{align}
Using the Tx ASE metric in \eqref{eq:tasedef}, the \emph{transmit NEE (Tx NEE)} is given as follows
\vspace{-0.2 cm}
\begin{align}
\label{eq:txnee}
\xi_t(\widetilde{\gamma}, \bm \lambda_b) & = \frac{B \, \mathcal{S}_t(\widetilde{\gamma}, \bm \lambda_b) }{ P_{\text{nec}} (\widetilde{\gamma}, \bm \lambda_b) } \, .
\end{align}
\noindent Similarly, the \emph{receive NEE (Rx NEE)} is given by using the Rx ASE in \eqref{eq:rasedef} as 
\vspace{-0.1 cm}
\begin{align}
\label{eq:rxnee}
\xi_r(\widetilde{\gamma}, \bm \lambda_b) & = \frac{B \, \mathcal{S}_r(\widetilde{\gamma}, \bm \lambda_b) }{ P_{\text{nec}} (\widetilde{\gamma}, \bm \lambda_b) } \, .
\end{align}
%
%

%
%

\section{Coverage Probability Analysis}\label{sec:perfanalysis}
In this section, using the definition of conditional SIR, $\gamma_{k}^{\ncob}(r)$, we derive the conditional coverage probability, $p_{\text{cov}}(r,k,\nco,\widetilde{\gamma})$ (see \eqref{eq:covpdef}). We note that $p_{\text{cov}}(r,k,\nco,\widetilde{\gamma})$ depends on the distribution of both $P_{\text{comp}}^{\ncob}(r)$ and $I^{\ncob}(r)$. In the following, we first analyze the distribution of $P_{\text{comp}}^{\ncob}(r)$, and then using this distribution, derive an expression for $p_{\text{cov}}(r,k,\nco,\widetilde{\gamma})$. Following this, we also present the analytical expression of the coverage probability for the simplified special case of NLoS link only scenario. 
\vspace{-0.4 cm}
\subsection{Coverage Probability}
From \eqref{eq:sigdef}, it is clear that the distribution of the received signal power for a given main link BS in the $k$-th tier with channel $\nco$ and link distance $r$ (i.e., $P_{\text{comp}}^{\ncob}(r)$) is determined by not only the number of cooperating BSs, $N^{\ncob}(r)$, but also the individual link distances and channels of the BSs in the cooperation set. Note that even with a given main link, the above mentioned parameters are likely to vary from one user to another, and therefore in general it is difficult to characterize the exact distribution of $P_{\text{comp}}^{\ncob}(r)$. Hence, in this section, we first fix the BS cooperation set for the typical user and then attempt to analyze the distribution of $P_{\text{comp}}^{\ncob}(r)$. For instance, where the main link BS for the typical user is in the $k$-th tier with link distance $r$ and channel $\nco$, we assume that the number of other cooperating BSs in the $j$-th tier with channel $\nc$ is $n_j^{\ncb}$, i.e., from \eqref{eq:compsetsizej}, we have
\vspace{-0.4 cm}
\begin{align}
\label{eq:nobseset}
N_j^{(\nco)}(r) & = \left \{ \begin{array}{ll}
\nL_j + \nN_j, & \text{if } j \neq k\\
\nL_j + \nN_j+1, & \text{if } j = k
\end{array} \right. \, .
\end{align}
\noindent Clearly, the total number of cooperating BSs for the typical user is $N^{\ncob}(r) = \sum\limits_{j=1}^{K}N_j^{(\nco)}(r)= 1+ \sum\limits_{j=1}^{K} \sum\limits_{\nc \in \{ \tL, \tN \} } n_j^{\ncb}$. Without loss in generality, we next index these other associated BSs (excluding the main link BS) in each tier as follows. The $j$-th tier BSs with LoS channel are indexed in the range $[1,n_j^{(\tL)}]$, and the BSs with NLoS channel are indexed in the range $[n_j^{(\tL)}+1, n_j^{(\tL)}+n_j^{(\tN)}]$. Thus, a $j$-th tier BS with channel $\nc$ has a link distance $x_{j,i}$ to the typical user, where $i \in [1,n_j^{(\tL)}]$, if $\nc = \tL$, and $i \in [n_j^{(\tL)}+1, n_j^{(\tL)}+n_j^{(\tN)}]$, if $\nc = \tN$.

In the following proposition, we now show that the distribution of $P_{\text{comp}}^{\ncob}(r)$ can be approximated to Gamma distribution for a given set of cooperating BSs.

\begin{proposition}\label{deistp}
\normalfont For a given set of cooperating BSs, $P_{\text{comp}}^{\ncob}(r)$ can be approximately Gamma distributed with shape and scale parameters $\zeta(\bm \varrho)$, and $\beta(\bm \varrho)$, where
\begin{align}
\label{eq:parseq}
\bm \varrho & \Define (\bar{x}_1, \cdots, \bar{x}_K,r,\nco,k) \,
\end{align}
\noindent and $\bar{x}_j \Define (x_{j,1}, \ldots, x_{j, \nL_j}, x_{j, \nL_j+1}, \ldots, x_{j, \nL + \nN})$. Here, $\zeta(\bm \varrho)$ and $\beta(\bm \varrho)$ are respectively given by
\vspace{-0.4 cm}
\begin{align}
\label{eq:param0}
\zeta(\bm \varrho) & = \frac{(\mu^2 + \Omega)^2}{\Omega_1}, \text{ and } \beta(\bm \varrho) \, = \frac{\Omega_1}{\mu^2 + \Omega} \, .
\end{align}
\indent Here $\mu$ and $\Omega$ are defined as
\vspace{-0.25 cm}
\begin{align}
\label{eq:param2}
\mu & = \hspace{-0.1 cm} \tau_{\tL,1} \sum\limits_{j=1}^{K}\sum\limits_{i = 1}^{\nL_j} x_{j,i}^{-\frac{\aL}{2}} + \tau_{\nco,1} r^{-\frac{\aco}{2}} \, +   \tau_{\tN,1} \sum\limits_{j=1}^{K}\sum\limits_{l = \nL_j + 1}^{\nL_j+\nN_j} x_{j,l}^{-\frac{\aN}{2}}\\
\Omega & =  \chi_{\nco} r^{-\aco} \,  + \chi_{\tL} \sum\limits_{j=1}^{K} \sum\limits_{i = 1}^{\nL_j} x_{j,i}^{-\aL} + \chi_{\tN} \sum\limits_{j=1}^{K} \sum\limits_{l= \nL_j + 1}^{\nL_j+ \nN_j} x_{j,l}^{-\aN}
\label{eq:param3}
\end{align}
\noindent where $\tau_{\nc, w} \Define \frac{\Gamma(m^{\ncb} + \frac{1}{2}w)}{\Gamma(m^{\ncb}) (m^{\ncb})^{w/2}}$ ($w \in \Z^{+}$),\footnote[7]{$\Z^{+}$ denotes the set of positive integers.} and $\chi_{\nc} \Define 1 - \tau_{\nc,1}^2$ ($\nc \in \{\tL, \tN \}$). In \eqref{eq:param0}, $\Omega_1 \Define \xi - (\mu^2 + \Omega^2)^2$, where $\xi$ is given by \cite[eq.~$6$]{Filho}
\begin{align}
\label{eq:fourthmoment}
\nonumber \xi & = \sum\limits_{s_0=0}^{4} \sum\limits_{s_1 = 0}^{s_0} \cdots \hspace{-0.5 cm} \sum\limits_{s_{N^{\ncob}(r)-1} = 0}^{s_{N^{\ncob}(r)-2}} {4 \choose s_0} {s_0 \choose s_1} \cdots {s_{N^{\ncob}(r)-2} \choose s_{N^{\ncob}(r)-1}} \hspace{-0.8 cm} \prod\limits_{\substack{j = 1,2, \ldots, K;\\ i = 1,2, \ldots, n_j^{(\tL)}; \\ l_{j,i}^{(\tL)} = \sum\limits_{t=1}^{j-1} n_t^{(\tL)} + n_t^{(\tN)} + i }} \hspace{0.2 cm} \prod\limits_{\substack{q = 1,2, \ldots, K;\\ v = 1,2, \ldots, n_q^{(\tN)}; \\ l_{q,v}^{(\tN)} = \sum\limits_{t=1}^{q-1} n_t^{(\tL)} + n_t^{(\tN)} + v }} \,\\
& \hspace{2 cm} \times \, \tau_{\tL, s_{l_{j,i}^{(\tL)}} - s_{l_{j,i}^{(\tL)}-1}} x_{j,i}^{-\frac{\aL}{2} \left(s_{l_{j,i}^{(\tL)}} \, - \, s_{l_{j,i}^{(\tL)}-1} \right ) } \tau_{\tN, s_{l_{q,v}^{(\tN)}} - s_{l_{q,v}^{(\tN)}-1}} x_{q,v}^{-\frac{\aN}{2}  \left(s_{l_{q,v}^{(\tN)}} \, - \,  s_{l_{q,v}^{(\tN)}-1} \right) } \, .
\end{align}
\end{proposition}
\begin{IEEEproof}
Since $g_{j,i}^{\ncb} \sim \Gamma \left( m^{\ncb}, \frac{1}{m^{\ncb}} \right)$, from \eqref{eq:rxpower}, we observe that for a given $x_{j,i}$ and a given channel $\nc$, $\sqrt{P_j^{\ncb}(x_{j,i})}$ is Nakagami-\textit{m} distributed. Thus, from \eqref{eq:sigdef}, it is clear that $P_{\text{comp}}^{\ncob}(r)$ is the squared sum of independent Nakagami-\textit{m} random variables. In \cite{Filho}, it has been shown that such a sum of independent Nakagami-\textit{m} random variables with different parameters can be approximated with sufficient accuracy to a Nakagami-\textit{m} distributed random variable, whose parameters can be computed from the moments of the component random variables. Using this principle, from \eqref{eq:sigdef}, we infer that $\sqrt{P_{\text{comp}}^{\ncob}(r)}$ is also approximately Nakagami-\textit{m} distributed, whose mean and variances are given by \eqref{eq:param2} and \eqref{eq:param3}, respectively. Clearly, the distribution of $P_{\text{comp}}^{\ncob}(r)$ can therefore be approximated to be a Gamma distribution whose parameters are given by \eqref{eq:param0}. 
\end{IEEEproof}
\begin{remark}
	\label{remarksigP}
\normalfont {\emph{(Impact of $\lambda_b$ on received signal power distribution)} From \eqref{eq:param0}-\eqref{eq:fourthmoment}, it is clear that both $\zeta(\bm \varrho)$ and $\beta(\bm \varrho)$ are depending on the link distances of the cooperating BSs. From the expressions of $\mu$ and $\Omega$, we also observe that the actual dependence of $\zeta(\bm \varrho)$ on the link distances is negligibly small since both the numerator and denominator in its expression have the same dependence on the link distances. For instance, if we consider the special scenario, where all the LoS and NLoS cooperating BSs are in the maximum distance away from the typical user (i.e., the worst CoMP set scenario), we have $x_{j,i} = \eta_{j,k}^{-\frac{1}{\alpha_{\tL} } } r_j^{(\tL)}$ for any LoS cooperating BS, and $x_{j,i+n_j^{(\tL)}} = \eta_{j,k}^{-\frac{1}{\alpha_{\tN} } }r_j^{(\tN)}$ for any NLoS cooperating BS, where $r_j^{\ncb} = \left( \frac{\varsigma_j}{\varsigma_k} \right)^{\frac{1}{\ack} } r^{\frac{\aco}{\ack}}$. Even with this worst case scenario, $\mu$, $\Omega$, and $\Omega_1$ are respectively proportional to $r^{-\aco/2}$, $r^{-\aco}$, and $r^{-2\aco}$. This shows that $\zeta(\bm \varrho)$ is independent of $r$ (the main link distance), while $\beta(\bm \varrho)$ is proportional to $ r^{-\aco}$. In short, $\beta(\bm \varrho)$ increases as the main link distance $r$ decreases, i.e., as the total BS density increases. An increase in $\beta(\bm \varrho)$ also causes a larger deviation in the received signal power from its mean, which in turn increases the probability of outage.}
\end{remark}
From the definition of the conditional SIR for a given main link BS, in the following, we can now derive $p_{\text{cov}}(r,k,\nco,\widetilde{\gamma})$, using the approximated received signal power distribution in Proposition~\ref{deistp}, where the typical user has its main link BS in the $k$-th tier with link distance $r$ and channel $\nco$.
\begin{theorem}\label{trm:condcovpc}
\normalfont The conditional coverage probability for a given main link BS in the $k$-th tier with link distance $r$ and channel $\nco$ can be expressed as a function of $\widetilde{\gamma}$ as 
\vspace{-0.2 cm}
\begin{align}
\label{eq:condcovpdef}
\nonumber & p_{\text{cov}}(r,k,\nco,\widetilde{\gamma}) \, = \sum\limits_{\nL_1 = 0}^{\infty} \sum\limits_{\nN_1 = 0}^{\infty} \cdots \sum\limits_{\nL_K = 0}^{\infty} \sum\limits_{\nN_K = 0}^{\infty} \left\{ \underbrace{\prod\limits_{w=1}^{K} \prod\limits_{\nc \in \{ \tL, \tN \} }  \frac{\left(\Lambda_w^{\ncb} \right)^{n_w^{\ncb}} }{n_w^{\ncb}!} e^{-\Lambda_w^{\ncb}}}_{\Define \, f_1(\bm \lambda_b)} \right \} \underbrace{\int_{r_{j}^{\ncp}}^{R_{j}^{\ncp}}}_{\substack{{j} = 1,2, \ldots, K; \, c^{\prime} \in \{ \tL, \tN \}; \\ \text{total} \, \sum_{j=1}^{K} (n_{j}^{(\tL)} + n_{j}^{(\tN)}) \\ \text{integrals} } } \\
& \hspace{-1 cm}  \times \, \underbrace{\prod\limits_{j=1}^{K} \, \frac{(2 \pi \lambda_{b,j})^{n_j^{(\tL)} + n_j^{(\tN)} }}{ (\Lambda_j^{ (\tL) })^{n_j^{(\tL)}} (\Lambda_j^{ (\tN) })^{n_j^{(\tN)}} } \,  \prod\limits_{i=1}^{\nL_j}\prod\limits_{l=1}^{\nN_j}      \, x_{j,i} \, x_{j,l+\nL_j} p^{(\tL)}(x_{j,i}) \, p^{(\tN)}(x_{j,l+\nL_j})}_{\Define f_2(\bm \lambda_b)} \,  \, p_{\text{cov}}(\widetilde{\gamma}, \bm \varrho) \hspace{-0.6 cm} \underbrace{dx_{j, i} dx_{j, l + n_{j}^{(\tL)} } }_{ \substack{i = 1,2, \cdots, n_{j}^{(\tL)}; l = 1,2, \cdots, n_{j}^{(\tN)};\\ j = 1,2, \ldots, K} }    
\end{align}
\noindent where $\bm \varrho = \{\bar{x}_1, \cdots, \bar{x}_K,r,\nco,k\}$, $\Lambda_j^{\ncb} \Define 2 \pi \lambda_{b,j} \int_{r_j^{\ncb}}^{R_j^{\ncb}} t \, p^{\ncb}(t) dt$, $r_j^{\ncb} = \theta_j^{\ncb}(r,\nco)$, $R_j^{\ncb} = \eta_{j,k}^{-1/\ack} \, r_j^{\ncb}$ ($\nc \in \{ \tL, \tN \}$). In \eqref{eq:condcovpdef}, $p_{\text{cov}}(\widetilde{\gamma}, \bm \varrho)$ is given by 
\vspace{-0.2 cm}
\begin{align}
\label{eq:condcondcovp2}
p_{\text{cov}}(\widetilde{\gamma}, \bm \varrho) & =  \sum\limits_{m=0}^{k_0(\bm \varrho) - 1} \frac{(-1)^m}{m!} \left(\frac{\widetilde{\gamma}}{\beta(\bm \varrho)}\right)^m \mathcal{L}_{I}^{(m)}\left(  \frac{\widetilde{\gamma}}{\beta(\bm \varrho)} \right)
\end{align}
\noindent where $k_0(\bm \varrho) \in \{ \lfloor \zeta(\bm \varrho) \rfloor, \lceil \zeta(\bm \varrho) \rceil \}$, and $\mathcal{L}_I^{(m)}(s)$ denotes the $m$-th order derivative of $\mathcal{L}_{I}(s) \, \Define \Eb{e^{-s \, I^{\ncob}(r)}}${, i.e.,}
\vspace{-0.5 cm}
{\begin{align}
	\label{eq:laplaceI}
	\mathcal{L}_{I}^{(m)}(s) & = \left \{\begin{array}{ll} e^{-2\pi\sum\limits_{j=1}^{K} \sum\limits_{\nc \in \{ \tL, \tN \} } \lambda_{b,j} \varDelta_{j,\nc}(s) }, & m=0\\
	-2\pi\sum\limits_{j=1}^{K} \lambda_{b,j} \sum\limits_{\nc \in \{ \tL, \tN \} } \sum\limits_{i=0}^{m-1} {m-1 \choose i} \mathcal{L}_I^{(i)}(s) \varDelta_{j,\nc}^{(m-i)}(s) , & m > 0 \end{array} \,\right. \, ,
\end{align}
\noindent where $\varDelta_{j,\nc}^{(m)}(s)$ is given by
\vspace{-0.4 cm}
\begin{align}
		\label{eq:laplaceI2}
		\varDelta_{j,\nc}^{(m)}(s) & = \left \{\begin{array}{ll} \int_{R_j^{\ncb}}^{\infty} \left(1 - \Eb{ e^{-s \varsigma_j g z^{-\ack}  } } \right) \, z \, p^{\ncb}(z) dz, & m=0\\
			(-1)^{m+1} \int_{R_j^{\ncb}}^{\infty} \varsigma_j^m \Eb{ g^m e^{-s \varsigma_j g z^{-\ack}  }   } z^{-m\ack + 1} p^{\ncb}(z)  dz , & m > 0 \end{array} \,\right. \, .
\end{align}}
\end{theorem}
\begin{IEEEproof}
See Appendix~\ref{app:condcovpp}. 
\end{IEEEproof}
\par The conditional coverage probability $p_{\text{cov}}(r,k,\nco,\widetilde{\gamma})$ computed in Theorem~\ref{trm:condcovpc} represents the coverage probability for the scenario, where the main link BS for the typical user is in the $k$-th tier and has a channel $\nco$ and link distance $r$. Substituting this expression along with \eqref{eq:mainlinktypeprob} in \eqref{eq:covpdef}, we can compute the overall coverage probability $\mathcal{P}_{\text{cov}}(\widetilde{\gamma},\bm \lambda_b)$ for the typical user in $K$-tier open-access UDN with the RRLP-based CoMP transmission in the downlink. 
\begin{corollary}
		\label{covcdiss} 
	\normalfont {The {conditional} coverage probability in (31) first increases with the total BS density $\lambda_b = \mathbf{1}^T\bm \lambda_b = \sum\limits_{j=1}^{K}\lambda_{b,j}$, and then decreases.} 
\end{corollary}
\begin{IEEEproof}
	{From \eqref{eq:laplaceI2}, we observe that $(-1)^{m-1}\varDelta_{j,\nc}^{(m)}(s)$ is a positive quantity, and can be approximately bounded as $$(-1)^{m-1}\varDelta_{j,\nc}^{(m)}(s) \geq p^{\ncb}(R_j^{\ncb})\frac{1}{\ack}s^{\frac{2}{\ack} - m} \varsigma_j^{\frac{2}{\ack}} \Eb{g^{\frac{2}{\ack}} \gamma\left(m-\frac{2}{\ack}, s\varsigma_j g \eta_{j,k} r^{-\aco}\right) }.$$ 
		Using this in \eqref{eq:laplaceI}, we can show that $(-1)^m\mathcal{L}_I^{(m)}(s) > 0$, for all $s > 0$. Using \eqref{eq:laplaceI} and \eqref{eq:laplaceI2} in \eqref{eq:condcondcovp2}, we observe that $p_{\text{cov}}(\widetilde{\gamma}, \bm \varrho)$ is the positive sum of the terms {with the} form $xe^{-x}$. Here $x$ is proportional to $\lambda_{b,j}$. Note that since any function with form $xe^{-x}$ has a single maximum point, all the terms in the summation in \eqref{eq:condcondcovp2} have their unique maxima point as well. Therefore, the overall sum, i.e., $p_{\text{cov}}(\widetilde{\gamma}, \bm \varrho)$ in \eqref{eq:condcondcovp2} also first increases with $\lambda_{b,j}$ ($j = 1,2,\ldots, K$), and then decreases. From \eqref{eq:condcovpdef}, we can note that $f_1(\bm \lambda_b)$ has also the form similar to $xe^{-x}$, and therefore first increases and then decreases with $\bm \lambda_b$. Additionally, $f_2(\bm \lambda_b)$ is independent of $\bm \lambda_b$. Therefore, we can conclude that $p_{\text{cov}}(r,k,\nco,\widetilde{\gamma})$ which is product of $p_{\text{cov}}(\widetilde{\gamma}, \bm \varrho)$, $f_1(\bm \lambda_b)$, and $f_2(\bm \lambda_b)$ would first increase with $\bm \lambda_b$, and then decrease.}
\end{IEEEproof}

{From Corollary~\ref{covcdiss} and \eqref{eq:covpdef}, we can also see that the overall coverage probability, $\mathcal{P}_{\text{cov}}(\widetilde{\gamma}, \bm \lambda_b)$, is expected to first increase and then decrease with the total BS density, $\lambda_b$. This can also be intuitively explained as follows. As BS density increases, the probability that a cooperating BS has LoS link to the user increases, which in turn increases the received signal power. In this regime, however, the increase in the received signal power is not significant due to longer link distances, and therefore the coverage probability increases with the BS density. However, for large BS density, the increase in received signal power less dominantly affect the coverage probability than the increase in interference power received from non-cooperating LoS BSs. Therefore, the overall coverage probability in \eqref{eq:covpdef} rapidly degrades with BS density. This conclusion is also supported in Fig.~\ref{fig:covt} and Fig.~\ref{fig:covp2t}.}
\vspace{-0.6 cm}
\subsection{Special Case: Performance Analysis for Rayleigh Fading Channel}\label{casestudy}
\begin{figure}[t!]
	\vspace{-0.7 cm}
	\begin{center}   
		{ 
			\includegraphics[width=0.55\columnwidth]{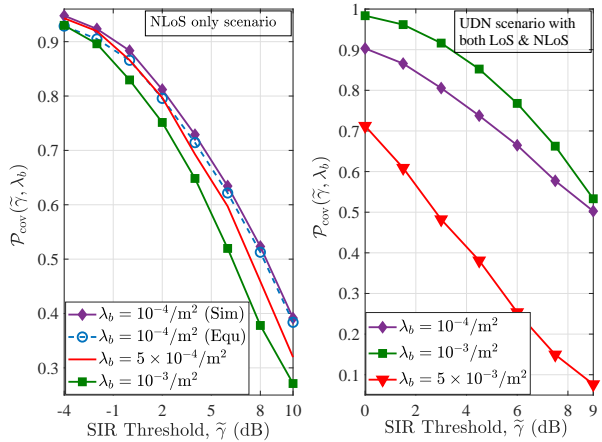}
		}
	\end{center}
	\caption{
		{$\mathcal{P}_{\text{cov}}(\widetilde{\gamma}, \lambda_b)$ as a function of $\lambda_b$ for the RRLP-based CoMP transmission with $1$-tier UDN (i.e., $K=1$), for $N_{\text{avg}} = 2$ in (1) the NLoS only scenario, and (2) the scenario with both LoS and NLoS.}	
	} \vspace{-0.5 cm}
	\label{fig:covt}
\end{figure}

{In the following, we provide the coverage probability of the $K$-tier UDN scenario, with the RRLP-based CoMP transmission, for a special case.} Here, we assume $p^{(\tL)}(x) = 0$, i.e., all BSs in the network are NLoS BSs, and the small scale fading is Rayleigh distributed (i.e., channel coefficients are independent and have complex Gaussian distribution with zero mean and unit variance). With this assumption, in the following, we derive an expression of the coverage probability.
\begin{lemma}\label{covpdefnlos}
\normalfont The coverage probability for the $K$-tier UDN with all NLoS BSs and Rayleigh fading channel is given by
%
%
\begin{align}
\label{eq:covpsc}
\nonumber & \mathcal{P}_{\text{cov}}(\widetilde{\gamma},\bm \lambda_b)  = \sum\limits_{k=1}^{K} \,2 \pi \, \lambda_{b,k} \int_{h_k}^{\infty} \,   r \exp \left \{-\pi \sum_{j=1}^{K} \lambda_{b,j}(\nu_{j,k}^2 \, r^2 - h_j^2) \right \}      \,  \sum\limits_{n_1 = 0}^{\infty} \cdots \sum\limits_{n_K=0}^{\infty} \left \{ \prod\limits_{l=1}^{K} \frac{(\Lambda_l)^{n_l}}{n_l !} \, e^{-\Lambda_l} \right \} \\
& \hspace{1 cm} \times \, \hspace{-1 cm} \underbrace{\int_{r}^{R_j}}_{\substack{j = 1,2, \ldots, K; \\ \text{total } \sum_{j=1}^{K} n_j \, \text{integrals}}}  \hspace{-1 cm} \left(\frac{2}{r^2} \right)^{n_1 + \cdots + n_K} \, \prod\limits_{i=1}^{n_j}\frac{x_{j,i}}{(\eta_{j,k}^{-\frac{2}{\alpha}} - 1)} \, \mathcal{L}_{I}\left(  {\widetilde{\gamma}}{\left(\varsigma_k r^{-\alpha} + \sum\limits_{j=1}^{K} \sum\limits_{i=1}^{n_j} \varsigma_j x_{j,i}^{-\alpha} \right)^{-1}} \right) \, d \bm x \, dr\, ,
\end{align}
\noindent where $\alpha > 2$ is the pathloss exponent, $\bm x \Define \{x_{1,1}, \ldots, x_{1,n_1}, \cdots, x_{K,1}, \ldots, x_{K, n_K}\}$ denotes the link distances of all cooperating BSs (excluding the main link BS), and $n_j$ is the number of such cooperating BSs in the $j$-th tier ($j = 1,2, \cdots, K$). Here, $\Lambda_j = \pi \lambda_{b,j}r^2 (\eta_{j,k}^{-\frac{2}{\alpha}} - 1)$, $R_j = \eta_{j,k}^{-\frac{1}{\alpha}} \, r$, and $\nu_{j,k} \Define \left(\frac{\varsigma_j }{\varsigma_k } \right)^{\frac{1}{\alpha}}$ ($k = 1,2, \cdots, K$). Finally, $\mathcal{L}_I(s)$ is the Laplace transform of the total interference power and is given by
%
%
\begin{align}
\label{eq:laplaceIsc}
\mathcal{L}_{I}(s) & =  \exp \left \{ -\pi \sum\limits_{j=1}^{K} \lambda_{b,j} \frac{\frac{2}{\alpha} \,s \,\varsigma_j \, R_j^{2-\alpha}}{1 - \frac{2}{\alpha}} \, \HGF{2}{1}{1,1-\frac{2}{\alpha};2-\frac{2}{\alpha};-s \, \varsigma_j\, R_j^{-\alpha}} \right \} \, ,
\end{align}
\noindent where $\HGF{2}{1}{a,b;c;z}$ represents the Gauss hypergeometric function.
\end{lemma}
\begin{IEEEproof}
	See Appendix~\ref{app:covpspcase}.
\end{IEEEproof}
{From \eqref{eq:covpsc}, it is clear that the conditional coverage probability is determined by $\mathcal{L}_I(s)$, where $s = {\widetilde{\gamma}}{\left(\varsigma_k r^{-\alpha} + \sum\limits_{j=1}^{K} \sum\limits_{i=1}^{n_j} \varsigma_j x_{j,i}^{-\alpha} \right)^{-1}}$. From \eqref{eq:laplaceIsc}, it is also observed that for any value of $s$, $\mathcal{L}_I(s)$ monotonically decreases with $\lambda_b$. In short, for the special case, the coverage probability is expected to decrease, as the total BS density increases. This conclusion is also supported from Fig.~\ref{fig:covt}. This essentially proves that the conventional scenario with all NLoS BSs and Rayleigh fading channel cannot completely characterize the performance in UDN channels.}

\begin{table}[!t]
\caption{UDN System Parameters \cite{IMT2020guide,IMT2020propagate} \label{table:paramlist}} 
\vspace{-0.7 cm}
\begin{center}
\rowcolors{2}
{cyan!15!}{}
\renewcommand{\arraystretch}{1.0}
\begin{tabular}{l | p{4 cm} | l || l | p{4 cm} | l }
\hline 
 {\bf Parameters} & {\bf Descriptions} & {\bf Values} &  {\bf Parameters} & {\bf Descriptions} & {\bf Values}   \\
\hline 
\hspace{0.15cm}$\varsigma_{1}$ & Downlink Tx. power (1st tier)
			& \hspace{0.12cm}$44$ dBm  &
			\hspace{0.15cm}$\varsigma_{2}$  & Downlink Tx. power (2nd ier)
			& \hspace{0.12cm}$33$ dBm \\ 
\hspace{0.15cm}$h_{b,1}$  & BS antenna height (1st tier)
						& \hspace{0.12cm}$25$ m   &
\hspace{0.15cm}$h_{b,2}$   & BS antenna height (2nd tier)
			& \hspace{0.12cm}$10$ m   \\ 
\hspace{0.15cm}$h_{u}$  & User antenna height
			& \hspace{0.12cm}$1.5$ m   &
			\hspace{0.15cm}$\widetilde{\gamma}$  & SIR Threshold
						& \hspace{0.12cm}$0$ dB   \\ 
\hspace{0.15cm}$\Upsilon$  & Average number of buildings/ unit area
			& \hspace{0.12cm}$300$/km\textsuperscript{2} &
			\hspace{0.15cm}$\varepsilon$  & Fraction of network area covered by buildings
						& \hspace{0.12cm}$0.5$ \\ 
\hspace{0.15cm} $\rho$    & Average building height
			& \hspace{0.12cm}$20$ m  &
\hspace{0.15cm}   & 
			& \hspace{0.12cm}  \\ 
\hline
\end{tabular} 
\end{center}
\end{table}%
%
%


\subsection{Evaluation of ASE Metrics}\label{evaluatease}
From Section~\ref{metrics}, we note that the Tx ASE metric requires computation of the average fractional contribution of each tier of BSs in the SE of a user (see \eqref{eq:tasedef}-\eqref{eq:txaseredef}). Therefore it requires numerical averaging of the per-link SE expression in \eqref{eq:txase}. 
\par On the other hand, however, the Rx ASE metric requires evaluation of the per-user SE (see \eqref{eq:rasedef}), and therefore can be evaluated from the coverage probability. From the definition of per-user SE, $\mathcal{R}_u(\widetilde{\gamma}, \bm \lambda_b)$, in \eqref{eq:puse}, we have
\vspace{-0.2 cm}
\begin{align}
\label{eq:casepcovdef}
\nonumber \mathcal{R}_u(\widetilde{\gamma}, \bm \lambda_b) & = \sum\limits_{k=1}^{K} \sum\limits_{\nco \in \{ \tL, \tN \} } \mathcal{A}_{k}^{\ncob} \int_{h_k}^{\infty} \, f_{X_{k,o}}^{\ncob}(r) \, \Eb{\log_2(1 + \gamma_{k}^{\ncob}(r)) \1{\gamma_{k}^{\ncob}(r) \geq \widetilde{\gamma}} } dr\\
& \mya \sum\limits_{k=1}^{K} \sum\limits_{\nco \in \{ \tL, \tN \} } \mathcal{A}_{k}^{\ncob} \int_{h_k}^{\infty} \, f_{X_{k,o}}^{\ncob}(r) \, \int_{\widetilde{\gamma}}^{\infty} \log_2(1 + t) f_{\gamma_{k}^{\ncob}(r)}(t) dt \, dr
\end{align}
\noindent where step $(a)$ follows from the fact that $f_{\gamma_{k}^{\ncob}(r)}(.)$ is the pdf of the conditional SIR $\gamma_{k}^{\ncob}(r)$, and is given by $f_{\gamma_{k}^{\ncob}(r)} (\gamma)\Define \frac{\partial}{\partial \gamma} \left [1 - p_{\text{cov}}(r,k,\nco,\gamma)\right]$, whereas $p_{\text{cov}}(r,k,\nco,{\gamma})$ is computed using Theorem~\ref{trm:condcovpc}.

%
%
%


\section{Simulations and Discussions}\label{sec:results}
In this section, we first validate our analysis of the coverage probability expression in Section~\ref{sec:perfanalysis}, and then investigate the performance of the RRLP-based CoMP transmission in multi-tier UDN in a dense urban scenario. For the simulations in the dense urban scenario, we consider a $K=2$-tier UDN network, with an area of $10$ sq.km, consisting of macro and micro BSs. The system parameter values are considered according to the recommendations of ITU \cite{IMT2020guide}, and are listed in Table~\ref{table:paramlist}. We also assume that the pathloss exponents for LoS and NLoS links are fixed and are respectively given by $\aL = 2.5$ and $\aN = 3.5$. The Nakagami-m parameter for channel fading gain is taken to be $\mL = 10$ and $\mN = 1$ for LoS and NLoS links respectively. Throughout this section, for simulations, we assume that the user density in the network is constant and equal to $\lambda_u = 3 \times 10^{-3}$/m\textsuperscript{2}. \cite{Niejin}
%
%
%

%
%
\subsection{Validation of Analytical Result for the Special case in Section~\ref{casestudy} }
In Fig.~\ref{fig:covt}, we have plotted the coverage probability for the NLoS channel only scenario as a function of SIR threshold $\widetilde{\gamma}$ for different values of the BS density in a $K=1$-tier network, i.e., $\lambda_{b,1} = \lambda_b$. For this simulation, we assume $N_{\text{avg}} = 2$, i.e., on average $2$ BSs associate with a user. We also assume $\aN = \alpha = 4$, and the transmission power parameters are adopted from Table~II. For comparison, we also numerically evaluate the expression in \eqref{eq:covpsc}, and plot in Fig.~\ref{fig:covt}. It is observed that there is negligible difference between these two plots, which verifies our analysis.

\begin{figure}[t!]
	\vspace{-0.8 cm}
	\begin{center}   
		{ 
			\includegraphics[width=0.55\columnwidth]{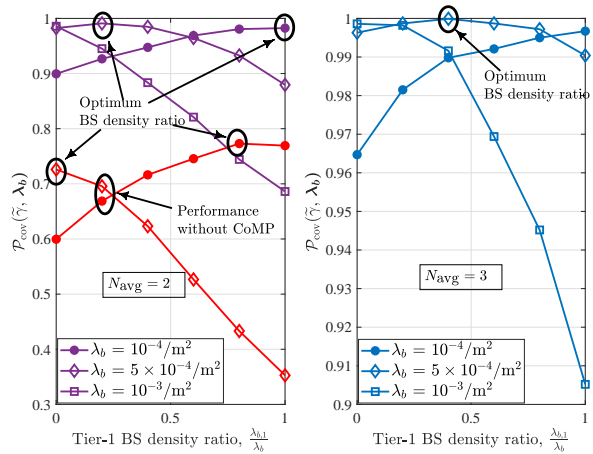}
		}
	\end{center}
	\caption{
		$\mathcal{P}_{\text{cov}}(\widetilde{\gamma}, \bm \lambda_b)$ as a function of $\frac{\lambda_{b,1}}{\lambda_b}$ with $N_{\text{avg}} = 2, 3$.	
	} \vspace{-0.5 cm}
	\label{fig:optrat}
\end{figure}

We also observe that as the BS density increases, the coverage probability monotonically decreases in the NLoS channel only scenario. However, in the UDN environment this does not hold true. Since in the UDN environment both the LoS and NLoS channels appear, with increasing BS density, the probability of LoS BSs associating with the typical user increases, which in turn improves the overage performance (see the plots for UDN scenario with LoS/NLoS channels in Fig.~\ref{fig:covt}). However, beyond a critical BS density, the increase in interference power due to more interfering LoS BSs begin to dominate and the coverage performance begins to degrade. This shows that for any given SIR threshold, there exists an optimum BS density which maximizes the coverage probability of the UDN with CoMP transmission. Note that this conclusion also holds true for the multi-tier UDN scenario (see Fig.~\ref{fig:optrat}).

\subsection{BS density ratio vs. Coverage Probability}
In Fig.~\ref{fig:optrat}, we first plot $\mathcal{P}_{\text{cov}}(\widetilde{\gamma}, \bm \lambda_b)$ as a function of the ratio of 1st tier BS density to the total BS density, i.e., $\frac{\lambda_{b,1}}{\lambda_b}$, ($\lambda_b = \bm{1}^T\bm \lambda_b = \sum\limits_{j=1}^{K} \lambda_{b,j}$) for different values of $\lambda_b = 10^{-4}$/m\textsuperscript{2}, $5 \times 10^{-4}$/m\textsuperscript{2}, and $10^{-3}$/m\textsuperscript{2} respectively. For this plot, we consider the average number of cooperating BSs, $N_{\text{avg}}$ fixed. We observe that for a given $N_{\text{avg}}$ and $\lambda_b$, $\mathcal{P}_{\text{cov}}(\widetilde{\gamma}, \bm \lambda_b)$ first increases with $\frac{\lambda_{b,1}}{\lambda_b}$, and beyond a critical value of this ratio, it begins to decrease. This is due to the fact that when $\lambda_b$ is small, the probability of a user associating with a 1st tier BS increases with 1st tier BS density (since both the transmission power and LoS probability are higher for 1st tier BSs). However, beyond a critical value of the BS density ratio, this increase in 1st tier BS density significantly increases the total received interference power, which in turn degrades the received SIR, and therefore $\mathcal{P}_{\text{cov}}(\widetilde{\gamma}, \bm \lambda_b)$ decreases. We also observe that this critical $\frac{\lambda_{b,1}}{\lambda_b}$, that maximizes $\mathcal{P}_{\text{cov}}(\widetilde{\gamma}, \bm \lambda_b)$, decreases with $\lambda_b$. In other words, at high BS densities, it is optimum to have all cooperating BSs in the 2nd tier. This critical $\frac{\lambda_{b,1}}{\lambda_b}$ is also observed to increase with $N_{\text{avg}}$ (see the plots for $\lambda_b = 5 \times 10^{-4}$/m\textsuperscript{2}). This is due to the fact that increasing $N_{\text{avg}}$ tends to mitigate more interference power, thereby allowing us to tolerate higher increase in 1st tier BS density for a given $\lambda_b$. This also shows that the RRLP based CoMP has higher interference mitigation capability when compared to the conventional transmission without CoMP.

\begin{figure}[t!]
    \vspace{-0.7 cm}
    \begin{center}   
    { 
	 \includegraphics[width=0.5\columnwidth]{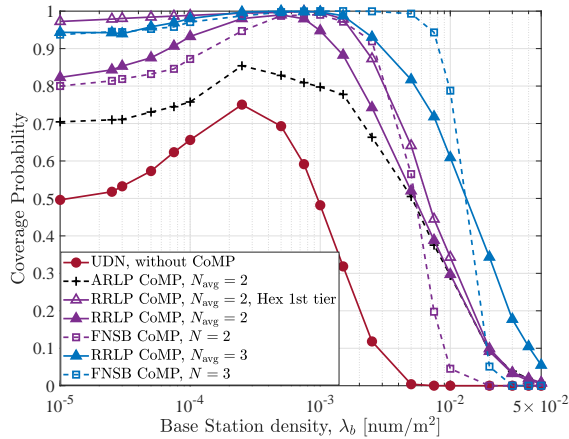}
    }
    \end{center}
    \caption{
    		{$\mathcal{S}_{t}(\widetilde{\gamma}, \bm \lambda_b)$ as a function of $\lambda_b$ in 2-tier UDN for the following transmission scenarios: (a) Without CoMP; (b) FNSB-based CoMP (in \cite{Nigam}); (c) ARLP-based CoMP (in \cite{Niejin}) and (d) RRLP-based CoMP.	}
		 } \vspace{-0.5 cm}
   \label{fig:covp2t}
\end{figure}

\vspace{-0.6 cm}

\subsection{Impact of Total BS density}
\begin{figure}[t!]
	\vspace{-0.2 cm}
	\begin{center}   
		{ 
			\includegraphics[width=0.49\columnwidth]{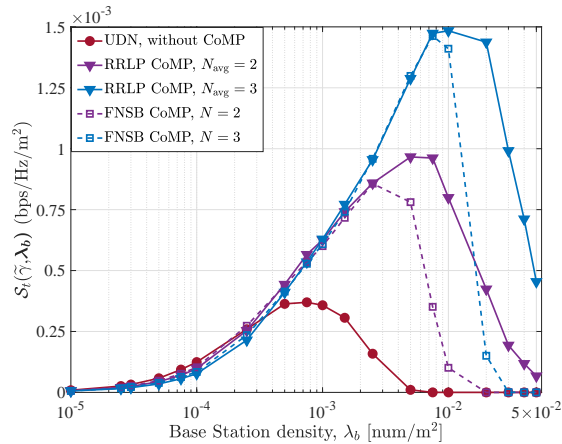}
		}
	\end{center}
	\caption{
		{$\mathcal{S}_{t}(\widetilde{\gamma}, \bm \lambda_b)$ as a function of $\lambda_b$ in 2-tier UDN for the following transmission scenarios: (a) Without CoMP; (b) FNSB-based CoMP (in \cite{Nigam}); and (c) RRLP-based CoMP.	}
	} \vspace{-0.5 cm}
	\label{fig:tase2t}
\end{figure}

\begin{figure}[t!]
	\vspace{-0.3cm}
	\begin{center}   
		{ 
			\includegraphics[width=0.49\columnwidth]{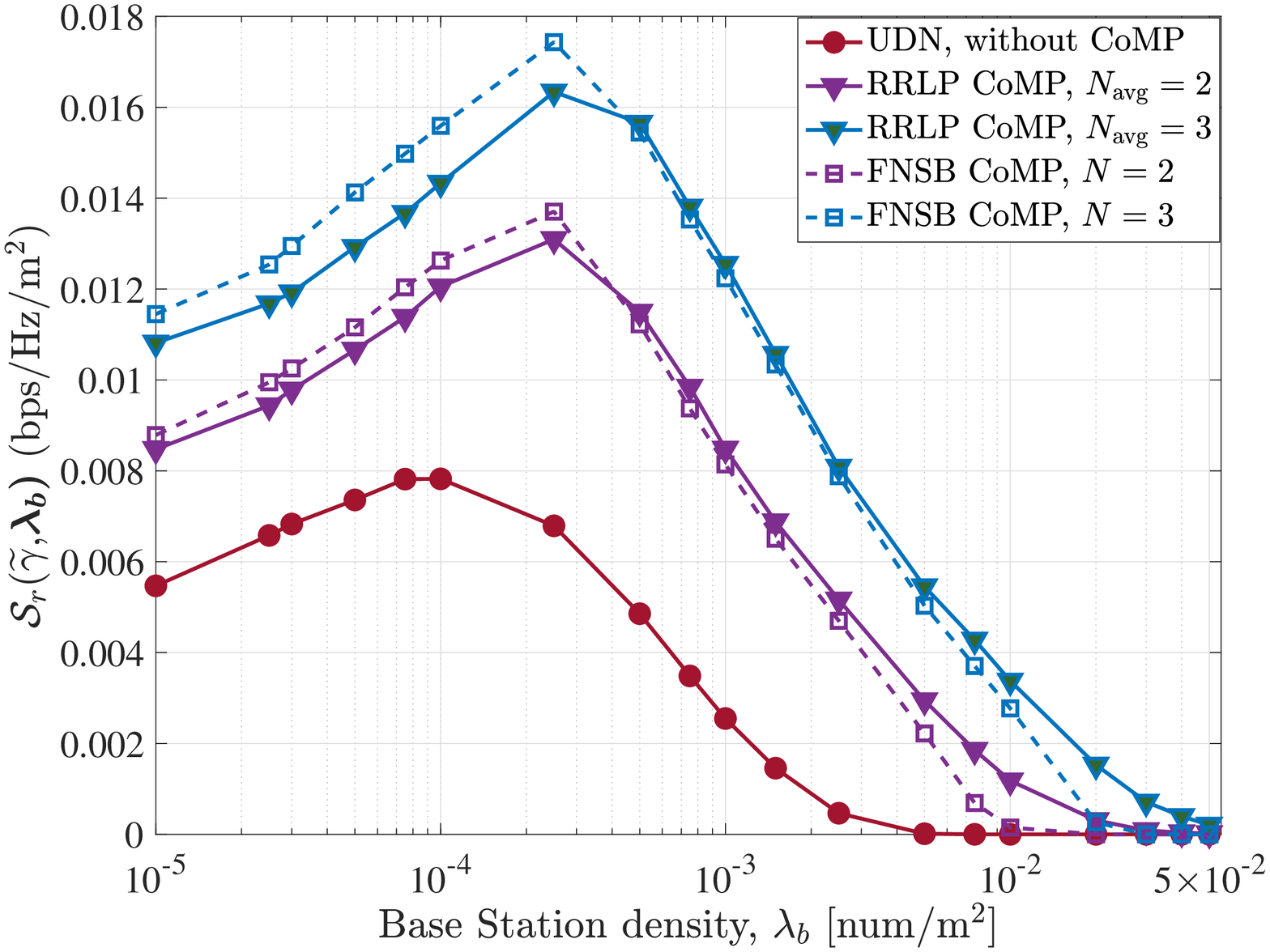}
		}
	\end{center}
	\caption{
		{$\mathcal{S}_{r}(\widetilde{\gamma}, \bm \lambda_b)$ as a function of $\lambda_b$ in 2-tier UDN for the following transmission scenarios: (a) Without CoMP; (b) FNSB-based CoMP (in \cite{Nigam}); and (c) RRLP-based CoMP.	}
	} \vspace{-0.5 cm}
	\label{fig:rase2t}
\end{figure}

Next in Fig.~\ref{fig:covp2t}, keeping $\frac{\lambda_{b,1}}{\lambda_b} = 0.2$ fixed,\footnote[8]{The $\frac{\lambda_{b,1}}{\lambda_b}$ ratio is chosen to be $0.2$, since it is observed to maximize the coverage probability in the network for a conventional moderate total BS density (in the order of $10^{-4}/$m\textsuperscript{2}).} we plot $\mathcal{P}_{\text{cov}}(\widetilde{\gamma}, \bm \lambda_b)$ as a function of $\lambda_b$ for $N_{\text{avg}} = 2$ and also for $N_{\text{avg}} = 3$, and compare with the traditional non-CoMP transmission performance \cite{Cholee, Kimlee}. This comparison shows significant performance improvement with CoMP, even for small BS density regime. For this scenario, we also plot the $\mathcal{S}_t(\widetilde{\gamma}, \bm \lambda_b)$ and $\mathcal{S}_r(\widetilde{\gamma}, \bm \lambda_b)$ metrics (see Fig.~\ref{fig:tase2t} and Fig.~\ref{fig:rase2t} respectively). It is observed that both the coverage probability and ASE metrics first improve with $\lambda_b$, and beyond a critical $\lambda_b$, they begin to decrease. This is due to the fact that in the small $\lambda_b$ regime, the probability that the cooperating BSs would have LoS links increases with the BS density, $\lambda_b$. This in turn significantly improves the received signal power. On the other hand, in the large $\lambda_b$ regime, the increase in the total interference power due to the higher number of LoS interference links dominates the SIR, and therefore, the overall performance degrades with $\lambda_b$.

{We also compare the performance of the RRLP-based CoMP scheme with existing other schemes such as the ARLP-based CoMP scheme \cite{Niejin}, of which CoMP set includes BSs with higher ARLP than a certain threshold, 
	and the fixed $N$-strongest BS (FNSB)-based CoMP scheme \cite{Nigam} (see Fig.~\ref{fig:covp2t}).} {It is observed that in the small BS density regime, the ARLP-based CoMP scheme performs worse, while the FNSB-based CoMP scheme performs better than the RRLP-based CoMP scheme. 
	When $\lambda_b$ is sufficiently large, the ARLP values can be large; so, for a fixed $N_{\text{avg}}$, the RRLP-based and ARLP-based CoMP schemes would have similar CoMP set size mostly. Therefore, in the large $\lambda_b$ regime, the performances of these two schemes track each other very closely. Additionally, the probability of having a larger CoMP set than $N_{\text{avg}}$ being high, which is more than enough to compensate the cases with the CoMP set size less than $N_{\text{avg}}$. Therefore, the performance of these two schemes exceeds that of the FNSB-based CoMP scheme.} Similar performance trend is also observed for both the Tx ASE and Rx ASE metrics. For instance, with the Tx ASE metric, $\mathcal{S}_t(\widetilde{\gamma}, \bm \lambda_b)$, for $\lambda_b = 5 \times 10^{-3}$/m\textsuperscript{2}, and $N_{\text{avg}} = 2$, the RRLP-based CoMP scheme out-performs the FNSB-based CoMP scheme by almost $24\%$.

{\emph{Comparison with Actual BS Deployment}: In Fig.~\ref{fig:covp2t}, we also plot the coverage probability of the RRLP-based CoMP scheme, where the 1st tier BSs are deployed in a hexagonal grid (as conventional macro BSs), while the distribution of the second tier BSs follows a PPP (marked by `Hex 1st tier' in Fig.~\ref{fig:covp2t}). The 1st tier BSs are assumed to be located at the center of hexagons, with a single sector and omni-directional beam pattern. This BS location distribution is adopted from \cite{Nasrijaziri}. In conventional HetNets, it is shown that the hexagonal grid-based deployment of BSs can provide higher performance than the PPP-based deployment \cite{Brown}, we can also see that when the RRPL-based CoMP scheme is applied in UDN, the hexagonal grid-based deployment of 1st tier BSs provides higher coverage probability, but the performance gap becomes smaller at higher BS densities.This means our analysis with PPP-based deployment of 1st tier BSs can also give a useful guideline in UDN, even for the hexagonal grid-based deployment case.}

\vspace{-0.4 cm}

\begin{table}[!t]
\caption{Communication \& Power Consumption Parameters ($j = 1,2$) \cite{HoydisEmil, Nakoh}} 
\label{table:commparam}
\vspace{-0.5 cm}
\begin{center}
\rowcolors{2}
{cyan!15!}{}
\renewcommand{\arraystretch}{1.0}
\begin{tabular}{p{1.4 cm} | p{4 cm} | l || p{1.4 cm} | p{4 cm} | l }
\hline 
 {\bf Parameters}  & {\bf Descriptions} & {\bf Values} &  {\bf Parameters} & {\bf Descriptions} & {\bf Values}   \\
\hline 
\hspace{0.15cm}$P_{\text{rf},j}^{(b)}$  & Antenna power at BS 
			& \hspace{0.12cm}$1$ Watt  &
\hspace{0.15cm}$P_{\text{rf}}^{(u)}$  & Antenna power at user
			& \hspace{0.12cm}$0.01$ Watt \\ 
			\hspace{0.15cm}$P_{\text{fix},j}$  & Fixed power consumed at BS 
						& \hspace{0.12cm}$18$ Watt   &
\hspace{0.15cm}$P_{\text{rate}}$   & Rate dependent power 
			& \hspace{0.12cm}$0.8$ $\frac{\text{Watt}}{\text{Gbits/sec}}$   \\ 
\hspace{0.15cm}$B_{\text{c}}$  & Coherence bandwidth
			& \hspace{0.12cm}$200$ kHz   &
			\hspace{0.15cm}$T_{\text{c}}$  & Coherence Time
						& \hspace{0.12cm}$1$ ms   \\ 
\hspace{0.15cm}$\frac{1}{\kappa_j}$ & Power efficiency of PA 
			& \hspace{0.12cm}$0.39$ &
			\hspace{0.15cm}$L_{b,j}$  & Computational efficiency of BS 
						& \hspace{0.12cm}$12.8$ $\frac{\text{Gflops}}{\text{Watt}}$ \\ 
\hspace{0.15cm}$C_0$   &  Duration of coherence block
			& \hspace{0.12cm}$B_{\text{c}} \, T_{\text{c}}$  & 
			\hspace{0.15cm}$B$ & Communication Bandwidth 
									& \hspace{0.12cm}$20$ MHz \\ 
\hline
\end{tabular} \vspace{-0.3cm}
\end{center}
\end{table}

\subsection{Network Energy Efficiency (NEE) Performance}
\begin{figure}[t!]
    \vspace{-0.7 cm}
    \begin{center}   
    { 
	 \includegraphics[width=0.49\columnwidth]{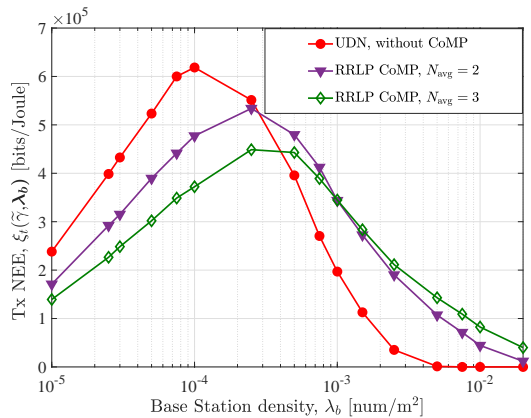}
    }
    \end{center}
    \caption{
    		{$\mathcal{\xi}_{t}(\widetilde{\gamma}, \bm \lambda_b)$ as a function of $\lambda_b$ in 2-tier UDN for the following transmission scenarios: (1) without CoMP; and (2) with RRLP-based CoMP, $N_{\text{avg}} = 2, 3$.}	
		 } \vspace{-0.5 cm}
   \label{fig:tnee2t}
\end{figure}

\begin{figure}[t!]
    \vspace{-0.2 cm}
    \begin{center}   
    { 
	 \includegraphics[width=0.49\columnwidth]{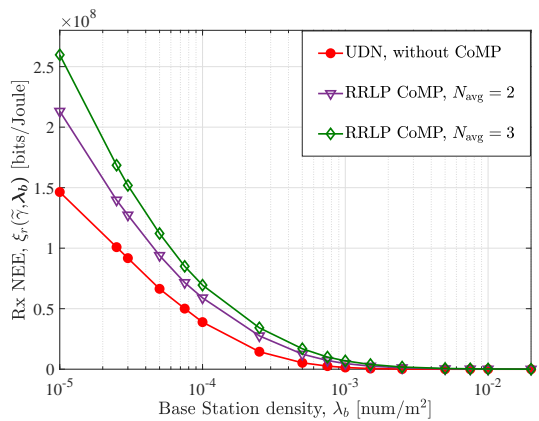}
    }
    \end{center}
    \caption{
    		{$\mathcal{\xi}_{r}(\widetilde{\gamma}, \bm \lambda_b)$ as a function of $\lambda_b$ in 2-tier UDN for the following transmission scenarios: (1) without CoMP; and (2) with RRLP-based CoMP, $N_{\text{avg}} = 2, 3$.	}
		 } \vspace{-0.5 cm}
   \label{fig:rnee2t}
\end{figure}

Finally, we analyze the NEE performance of our RRLP CoMP scheme as a function of $\lambda_b$, for a given $N_{\text{avg}}$. We assume the values of various transmission and circuit power consumption parameters as listed in Table~\ref{table:commparam}. To analyze the NEE performance, we plot both the Tx NEE, $\xi_t(\widetilde{\gamma}, \bm \lambda_b)$, and Rx NEE, $\xi_r(\widetilde{\gamma}, \bm \lambda_b)$ as functions of $\lambda_b$ (see Fig.~\ref{fig:tnee2t}, and Fig.~\ref{fig:rnee2t}). We also plot the  NEE corresponding to the scenario without CoMP transmission. It is observed that in the high $\lambda_b$ scenarios, our CoMP scheme out-performs the conventional non-CoMP transmission, in terms of achievable NEE (in contrast to the low $\lambda_b$ scenario). Furthermore, for a given $\lambda_b$ in this regime, we also observe that NEE improves with increasing $N_{\text{avg}}$. This shows that in UDN, our CoMP transmission strategy not only provides an improved performance in terms of both coverage probability and ASE, but also is more energy efficient when compared to the conventional transmission without CoMP. 

\vspace{-0.7 cm}


\section{Conclusion}\label{sec:conclusion}
In this paper, we introduce a relative received link power based CoMP transmission strategy for UDN. Considering the ARLP of the strongest BS to the typical user, we include the other BSs in the cooperation set when their ARLP exceed a certain ratio of the strongest BS's ARLP. For this CoMP scheme, we analyze the coverage probability, ASE and NEE in a $K$-tier UDN network, and we show that the coverage probability for this scheme first increases and then decreases with $\lambda_b$. From the simulation results, it is also observed that the RRLP-based CoMP scheme can outperform the FNSB-based CoMP scheme when the total BS density is high. Furthermore, we also show that the RRLP-based CoMP scheme is more energy efficient when compared to the conventional transmission scenario without CoMP in the high BS density regime. 

{To realize the CoMP transmission in practice, the support of the reliable backhaul links is essential as BSs need to exchange informations such as channel state, synchronization, and user data. Recently, the wireless backhaul is also considered as a more realistic implementation of backhaul in UDN. This opens several issues for future research such as efficient CoMP design with limited backhaul link capacity in UDN and the frequency resource management for access and backhaul links.}

%



\vspace{-0.6 cm}

\begin{appendix}

\subsection{Proof of Lemma~\ref{mainlinkpdfc}}\label{app:pdfcproof}
Note that the conditional cumulative distribution of $X_{k,o}^{\ncob}$ for a given main link BS in the $k$-th tier with channel $\nco$, is given by
\vspace{-0.4 cm}
\begin{align}
	\label{eq:probmainbs}
	\Pb{X_{k,o}^{\ncob} \leq r \Big| \nco, k} & = 1 - \Pb{X_{k,o}^{\ncob} \geq r \Big| \nco, k } \, =  1 - \frac{\Pb{X_{k,o}^{\ncob} \geq r , \nco, k}}{\mathcal{A}_{k}^{\ncob}} \,.
\end{align}
\noindent where $\mathcal{A}_{k}^{\ncob}$ is the tier association probability for the main link BS, and $\Pb{X_{k,o}^{\ncob} \geq r, \nco,k }$ denotes the probability that all BSs with channel $\nc \neq \nco$ in the $k$-th tier and BSs from other tiers do not have their ARLP higher than that of the designated main link BS.

Next we first define the void probability and contact distance distribution for the multi-tier UDN scenario, using which we can derive the expressions for $\mathcal{A}_{k}^{\ncob}$ and $\Pb{X_{k,o}^{\ncob} \geq r, \nco,k }$. Let us consider the $k$-th tier BSs in the network, with channel $\nc$ to the typical user, and their corresponding PPP is given by $\bm \Phi_{b, k}^{\ncb}$. Let $Y_k^{\ncb}$ denote the variable representing the horizontal distance of the nearest BS in this tier and let $X_k^{\ncb} = \sqrt{(Y_k^{\ncb})^2 + h_k^2}$ denote the corresponding link distance variable. Thus the void probability for the above mentioned BSs would be given by $V^{\ncb}_k(Y_k^{\ncb} = y) = \Pb{\left| \bm \Phi_{b, k}^{\ncb} \left( \mathcal{B}(o,y) \right) \right| = 0} = e^{-2\pi \lambda_{b,k} \, \int_{0}^{y} t \, p^{\ncb}(\sqrt{t^2+h_k^2}) \, dt}$.\footnote[9]{$\mathcal{B}(o,y)$ is a ball of radius $y$ with center at origin, and $\left| \bm \Phi_{b, k}^{\ncb} \left( \mathcal{B}(o,y) \right) \right|$ is the number of BSs from PPP $\bm \Phi_{b, k}^{\ncb}$ in $\mathcal{B}(o,y)$.} Correspondingly, the contact distance distribution can be computed as follows $f^{\prime}_{Y_{k}^{\ncb}}(y) \Define \frac{\partial}{\partial y} \left\{1 - V^{\ncb}_k(y)\right\} = 2\pi\lambda_{b,k} y p^{\ncb}(\sqrt{y^2+h_k^2})V^{\ncb}_k(y)$. Substituting $Y_k^{\ncb}$ with the link distance $X_k^{\ncb}$ in the expressions of void probability and contact distance distribution, we have
\vspace{-0.4 cm}
\begin{align}
	\label{eq:void}
	V^{\ncb}_k(X_k^{\ncb} = x) & = e^{-2\pi \lambda_{b,k} \int_{h_k}^{x} t \, p^{\ncb}(t) dt } \, , \text{and } \, f^{\prime}_{X_k^{\ncb}}(x) \, = 2 \pi \lambda_{b,k} x p^{\ncb}(x) \, V^{\ncb}_k(x) \, .
\end{align}
Now we derive an expression for $\mathcal{A}_{k}^{\ncob}$ as follows. From our discussion of the CoMP transmission strategy in Section~\ref{rrpcomp}, it is clear that the main link BS for the typical user is in the $k$-th tier, with channel $\nco$ if and only if $\bar{P}_k^{\ncob}(x_o) \, > \bar{P}_j^{\ncb}(x_{j,i})$, for all $(k,\nco) \neq (j,\nc)$ ($j = 1,2, \cdots, K$, and $\nc \in \{\tL, \tN \} $). Therefore, we have
\vspace{-0.2 cm}
\begin{align}
	\label{eq:linkassoprob}
	\nonumber \mathcal{A}_{k}^{\ncob} & = \int_{h_k}^{\infty}f^{\prime}_{X_{k,o}^{\ncob}}(x)\Pb{x_{j,i} > \theta_j^{\ncb}(x,\nco) , (k, \nco) \neq (j, \nc), \forall \, i \in \bm \phi_{b,j}^{\ncb} , j \in \{1,2,\ldots, K\} } \, dx\\
	& = \int_{h_k}^{\infty}f^{\prime}_{X_{k,o}^{\ncob}}(x) \, \prod\limits_{\stackrel{(k, \nco) \neq (j, \nc)}{\substack{j \in \{1,2,\ldots,K\};  \nc \in \{ \tL, \tN \}} }} V^{\ncb}_j \left(\theta_j^{\ncb}(x,\nco) \right) dx \, .
\end{align}
\noindent where $\theta_j^{\ncb}(x,\nco) = \left(\frac{\varsigma_j }{\varsigma_k } \right)^{\frac{1}{\ack}} \, x^{\frac{\aco}{\ack}}$. Finally, using \eqref{eq:void} in \eqref{eq:linkassoprob}, we obtain \eqref{eq:mainlinktypeprob}. Using \eqref{eq:void} in the definition of $\Pb{X_{k,o}^{\ncob} \geq r, \nco,k }$, we have
\vspace{-0.2 cm}
\begin{align}
	\label{eq:mainlinkprobtype}
	\nonumber \Pb{X_{k,o}^{\ncob} \geq r , \nco,k} & = \int_{r}^{\infty} f^{\prime}_{X_{k,o}^{\ncob}}(x) \, \Pb{\bar{P}^{\ncb}_j(x_{j,i}) < \bar{P}^{\ncob}_k(x) , (k, \nco) \neq (j, \nc), \forall \, i \in \bm \phi_{b,j}^{\ncb}} dx\\
	& = \int_{r}^{\infty} f^{\prime}_{X_{k,o}^{\ncob}}(x) \prod\limits_{\stackrel{(k, \nco) \neq (j, \nc)}{\substack{j \in \{1,2,\ldots,K\} ; \nc \in \{ \tL, \tN \}} }} V^{\ncb}_j \left( \theta_j^{\ncb}(x,\nco) \right) dx 
\end{align}
\noindent where $f^{\prime}_{X_{k,o}^{\ncob}}(x)$ and $V^{\ncb}_j (x)$ are defined in \eqref{eq:void}. Substituting \eqref{eq:mainlinkprobtype} in \eqref{eq:probmainbs} after using \eqref{eq:void}, and differentiating with respect to $r$, we get \eqref{eq:mainlinkpdf}.


\vspace{-0.6 cm}

\subsection{Proof of Lemma~\ref{avgbssetsize}}\label{app:avgbssetsizep}
\vspace{-0.3 cm}
The average number of cooperating BSs in the $j$-th tier for the typical user, whose main link BS is in the $k$-th tier with channel $\nco$, is given by
\vspace{-0.2 cm}
\begin{align}
	\label{eq:avgbsetj}
	\nonumber \bar{N}_j^{(\nco)}(x_{k,o}) &  \mya \, \Eb{\sum\limits_{ \nc \in \{ \tL, \tN \} } \sum\limits_{i \in \bm \phi_{b, j}^{\ncb} } \1{x_{j,i} \leq   \theta_{j}^{\ncb}(x_{k,o},\nco) \, \eta_{j,k}^{-\frac{1}{\ack}}} \Big | \nco, x_{k,o} }\\
	& \myb \, 2\pi \lambda_{b,j} \sum\limits_{\nc \in \{ \tL, \tN \} } \int_{h_j}^{\theta_j^{\ncb}(x_{k,o},\nco) \eta_{j,k}^{-\frac{1}{\ack}}} t \, p^{\ncb}(t) \, dt \, ,
\end{align}
\noindent where $(a)$ follows from \eqref{eq:compsetsizej}, $(b)$ follows from the Campbell's theorem \cite{Haenggi}, and $p^{\ncb}(x)$ is the probability of having channel environment $\nc$ for a link distance $x$. Clearly, the total average number of associated BSs for the typical user for the CoMP transmission scenario is given by 
\vspace{-0.2 cm}
\begin{align}
	\label{eq:avgbssizejn}
	N_{\text{avg}} & = \sum\limits_{k=1}^{K} \,\sum\limits_{ \nco \in \{ \tL, \tN \} } \, \mathcal{A}_{k}^{\ncob} \E_{x_{k,o}}\left [\sum\limits_{j=1}^{K}\bar{N}_j^{(\nco)}(x_{k,o})\right] \, ,
\end{align}
\noindent where $\mathcal{A}_{k}^{\ncob}$ is the probability that the typical user has its main link BS in the $k$-th tier with channel $\nco$. Substituting \eqref{eq:avgbsetj} in \eqref{eq:avgbssizejn}, and using \eqref{eq:mainlinkpdf}, we obtain \eqref{eq:avgbssize}.
%
%


\vspace{-0.6 cm}

\subsection{Proof of Theorem~\ref{trm:condcovpc}}\label{app:condcovpp}
Let us first consider the distribution of BSs from the $j$-th tier with channel $\nc$ to the typical user ($\nc \in \{ \tL, \tN \}$). The set of such BSs is given by $\bm \phi_{b,j}^{\ncb}$, and in the link distance $x$ to $x + dx$ ($dx \to 0$) from the typical user, the density of such BSs would be given by $2 \, \pi \lambda_{b,j} \, x \, p^{\ncb}(x) \, dx$. Any BS from this set will be included in the CoMP set, if its corresponding link distance lies in the following region $x \in [r_j^{\ncb}, R_j^{\ncb}]$, where $R_j^{\ncb}$ is obtained from the CoMP set formation criterion in \eqref{eq:compcondjk} and $r_j^{\ncb}$  is obtained from the fact that for any BS $i$ from the $j$-th tier, $\bar{P}_j^{\ncb}(x_{j,i}) < \bar{P}_k^{\ncob}(r)$, where $r$ is the main link distance. Clearly, $r_j^{\ncb} = \theta_j^{\ncb}(r,\nco)$, and $R_j^{\ncb} = \eta_{j,k}^{-{1}/{\ack}} r_j^{\ncb}$. 

Let $N_j^{\ncb}$ denote the number of cooperating BSs in the $j$-th tier ($j = 1,2 \ldots, K$) with channel $\nc$, excluding the main link BS. Since the BS distribution of the $j$-th tier follows a PPP, following the Slivnyak's theorem \cite{Mecke}, $N_j^{\ncb}$ can be described as a Poisson random variable, with mean $\Lambda_j^{\ncb} = 2 \pi \lambda_{b,j} \int_{r_j^{\ncb}}^{R_j^{\ncb}} t \, p^{\ncb}(t) \, dt$. Therefore, its probability mass function is given by $\Pb{N_j^{\ncb} = n_j^{\ncb}}  = \frac{(\Lambda_j^{\ncb})^{n_j^{\ncb}}}{n_j^{\ncb}!} \, e^{-\Lambda_j^{\ncb} } $. Now, denoting the vectors of link distances from the $j$-th tier cooperating BSs as $\bar{x}_j$, we represent the coverage probability for a given set of cooperating BSs as $p_{\text{cov}}(\widetilde{\gamma}, \bm \varrho)$, where $\bm \varrho = \{\bar{x}_1, \cdots, \bar{x}_K,r,\nco,k\}$. Clearly, the overall coverage probability for a given main link BS in the $k$-th tier with channel $\nco$ and link distance $r$ is given by
\vspace{-0.2 cm}
\begin{align}
\label{eq:condcovp1}
p_{\text{cov}}(r,k,\nco,\widetilde{\gamma}) & = \sum\limits_{\nL_1 = 0}^{\infty}\sum\limits_{\nN_1 = 0}^{\infty} \cdots \sum\limits_{\nL_K = 0}^{\infty}\sum\limits_{\nN_K = 0}^{\infty} \left \{ \prod\limits_{j=1}^{K} \prod\limits_{\nc \in \{ \tL, \tN \} } \Pb{N_j^{\ncb} = n_j^{\ncb}} \Eb{p_{\text{cov}}(\widetilde{\gamma}, \bm \varrho)} \right\} \, .
\end{align}
\noindent Here $\Eb{.}$ is taken over $\bm \varrho$, i.e.,
\vspace{-0.15 cm}
\begin{align}
\label{eq:condcovp2}
 & \hspace{-1 cm} \Eb{p_{\text{cov}}(\widetilde{\gamma}, \bm \varrho)}  = \hspace{-1.2 cm} \underbrace{\int_{r_j^{\ncb}}^{R_j^{\ncb}}}_{ \substack{ n_j^{\ncb} \, \text{times}; \\ j = 1,2,\ldots, K; \\ \text{total} \, \sum\limits_{j=1}^{K} ( n_j^{(\tL)} + n_j^{(\tN)} ) \, \text{integrals}    } } \hspace{-1.2 cm} \, \prod\limits_{j=1}^{K} \,   \prod\limits_{i=1}^{\nL_j}\prod\limits_{l=1}^{\nN_j}  \,  f_{X_{j,i}^{(\tL)}} \, (x_{j,i})  \, f_{X_{j,l}^{(\tN)}}(x_{j, l+\nL_j}) p_{\text{cov}}(\widetilde{\gamma}, \bm \varrho) \underbrace{dx_{j,i}}_{\substack{j = 1,2, \ldots, K; \\ i = 1,2, \ldots, n_j^{(\tL)}}} \underbrace{dx_{j,l+ n_j^{(\tL)} }}_{\substack{j = 1,2, \ldots, K; \\ l = 1,2, \ldots, n_j^{(\tN)}}} 
\end{align}
\noindent where $f_{X_{j,i}^{\ncb}}(.)$ denotes the pdf of $i$-th cooperating BS from the $j$-th tier with link type $\nc$, for the given main link BS. Not distinguishing among these cooperating BSs in terms of link distance, the conditional link distance distribution is given by
\vspace{-0.2 cm}
\begin{align}
\label{eq:condlinkpdf}
f_{X_{j,i}^{\ncb}}(x) & = \frac{\partial}{\partial x} \left\{\frac{1}{ \Lambda_j^{\ncb}} 2 \pi \lambda_{b,j}\int_{r_j^{\ncb}}^{x}t \, p^{\ncb}(t) \right\}  \, = \, \frac{2 \pi \lambda_{b,j}}{\Lambda_j^{\ncb}} x p^{\ncb}(x) .
\end{align}
\par {Using \eqref{eq:condcovp2} and \eqref{eq:condlinkpdf} in \eqref{eq:condcovp1}, we obtain \eqref{eq:condcovpdef}. Finally, for a given cooperation set of BSs, $p_{\text{cov}}(\widetilde{\gamma}, \bm \varrho)$ is given by
\begin{align}
\label{eq:compprob}
\nonumber p_{\text{cov}}(\widetilde{\gamma}, \bm \varrho) & \mya \Eb{\frac{\Gamma(\zeta(\bm \varrho), \frac{\widetilde{\gamma}}{\beta(\bm \varrho)})}{\Gamma(\zeta(\bm \varrho))}} \,  \mbox{\scriptsize $ {\gleq \atop {k_0(\bm \varrho) = \lceil \zeta(\bm \varrho) \rceil}}$ } \sum\limits_{m=0}^{k_0(\bm \varrho) - 1} \frac{(-1)^m}{m!} \,  \Eb{\left(\frac{\widetilde{\gamma}\, I^{\ncob}(r)}{\beta(\bm \varrho)}\right)^m e^{-\frac{\widetilde{\gamma}}{\beta(\bm \varrho)}I^{\ncob}(r)}}\\
& = \sum\limits_{m=0}^{k_0(\bm \varrho) - 1} \frac{(-1)^m}{m!} \left(\frac{\widetilde{\gamma}}{\beta(\bm \varrho)}\right)^m \, \mathcal{L}_{I}^{(m)}\left(  \frac{\widetilde{\gamma}}{\beta(\bm \varrho)} \right) \, .
\end{align}
\noindent where step $(a)$ follows from the fact that the received signal power is approximately Gamma distributed with parameters $\zeta(\bm \varrho)$ and $\beta(\bm \varrho)$ (see Proposition~\ref{deistp}). Here, $\mathcal{L}_{I}(s)$ is the Laplace transform of $I^{\ncob}(r)$. Using the definition of probability generating functional from \cite{Mecke} in the definition of $\mathcal{L}_{I}(s)$, we get \eqref{eq:laplaceI} and \eqref{eq:laplaceI2}, by using Leibniz integral rule \cite{Protter}.}
\vspace{-0.5 cm}

\subsection{Proof of Lemma~\ref{covpdefnlos}}\label{app:covpspcase}
From the coverage probability definition in \eqref{eq:covpdef}, we have
\vspace{-0.35 cm}
\begin{align}
\label{eq:covpsc0}
\mathcal{P}_{\text{cov}}(\widetilde{\gamma},\bm \lambda_b) & = \sum\limits_{k=1}^{K} \mathcal{A}_k \int_{h_k}^{\infty} f_{X_{k,o}}(r) p_{\text{cov}}(r,k,\widetilde{\gamma}) dr \, .
\end{align}
\noindent where $\mathcal{A}_k$ is the probability that the main link BS is in the $k$-th tier. The main link distance is denoted as $X_{k,o}$ and $f_{X_{k,o}}(r)$ is its pdf. Finally, $p_{\text{cov}}(r,k,\widetilde{\gamma})$ is the conditional coverage probability, corresponding to the main link BS in the $k$-th tier with main link distance $r$. 
Firstly, using Lemma~\ref{avgbssetsize}, from \eqref{eq:mainlinktypeprob} we have $\mathcal{A}_k = \int_{h_k}^{\infty} 2 \pi \lambda_{b,k} x e^{-\pi \sum_{j=1}^{K} \lambda_{b,j} (\nu_{j,k}^2 \, x^2 - h_j^2)} dx$, where $\nu_{j,k} \Define \left(\frac{\varsigma_j }{\varsigma_k } \right)^{\frac{1}{\alpha}}$ ($j = 1,2, \ldots, K$, and $k = 1,2, \ldots, K$). Next using \eqref{eq:mainlinkpdf} from Lemma~\ref{mainlinkpdfc}, we obtain the main link pdf $f_{X_{k,o}}(r)  = \frac{2 \pi \, \lambda_{b,k}}{\mathcal{A}_k} r e^{-\pi \sum_{j=1}^{K} \lambda_{b,j} (\nu_{j,k}^2 \, r^2 - h_j^2)}$. Substituting $\mathcal{A}_k$ and $f_{X_{k,o}}(r)$ in \eqref{eq:covpsc0}, we have
\vspace{-0.5 cm}
\begin{align}
\label{eq:revcovpsc1}
\mathcal{P}_{\text{cov}}(\widetilde{\gamma},\bm \lambda_b) & = \sum\limits_{k=1}^{K} 2 \pi \lambda_{b,k}  \int_{h_k}^{\infty} r e^{-\pi \sum_{j=1}^{K} \lambda_{b,j} (\nu_{j,k}^2 \, r^2 - h_j^2)} p_{\text{cov}}(r,k,\widetilde{\gamma}) dr \, .
\end{align}
\par Next we use Theorem~\ref{trm:condcovpc} to evaluate $p_{\text{cov}}(r,k,\widetilde{\gamma})$. This requires the conditional distance distribution of all cooperating BSs for the given main link BS in the $k$-th tier with link distance $r$. Using \eqref{eq:condlinkpdf}, we have $f_{X_{j,i}}(x) = \frac{2 \pi \lambda_{b,j} }{\Lambda_j} x$, where $\Lambda_j = \pi \lambda_{b,j}r^2 (\eta_{j,k}^{-\frac{2}{\alpha}} - 1)$. Furthermore, from Proposition~\ref{deistp}, we note that $P_{\text{comp}}(r)$ is $\Gamma(1, \beta(\bm \varrho))$ distributed, where $\beta(\bm \varrho) = \left(\varsigma_k r^{-\alpha} + \sum\limits_{j=1}^{K} \sum\limits_{i=1}^{n_j} \varsigma_j x_{j,i}^{-\alpha} \right)$. Substituting this expression of $f_{X_{j,i}}(x)$ and $\beta(\bm \varrho)$ in \eqref{eq:condcovpdef}, we obtain \eqref{eq:covpsc}, where $\mathcal{L}_{I}(s)$ is the Laplace transform of the total interference, and is computed by using \cite[eq.~$3.194$]{Ryzhik}, as in \eqref{eq:laplaceIsc}.

\end{appendix}

%
%
%
\ifCLASSOPTIONcaptionsoff
  \newpage
\fi

\vspace{-0.7 cm}



%

\bibliographystyle{IEEEtran}
\bibliography{IEEEabrvn,StringDefinitions,mybibn}

\end{document}